\documentclass[preprint,aps]{revtex4-1}
\usepackage{graphicx}
\usepackage{rotating}
\usepackage{amsmath}
\usepackage{epsfig}
\usepackage{bm}

\begin{document}
\topskip 20mm
\title{Boson peak in amorphous systems:  role of phonon mediated coupling of nano-clusters}
\author{Pragya Shukla}
\affiliation{Department of Physics, Indian Institute of Technology, Kharagpur-721302, India }
\date{\today}
\widetext

\begin{abstract}

Based on a description of an amorphous solid as a collection of coupled nanosize molecular clusters referred as basic blocks,  we analyse the statistical properties of its Hamiltonian. The information is then used to derive the ensemble averaged density of the vibrational states (non-phonon) which turns out to be a Gaussian in the bulk of the spectrum and an Airy function in the low frequency regime. A comparison with  experimental data for five glasses confirms validity of our theoretical predictions. 

\end{abstract}

\maketitle

\section{Introduction}
.

As indicated by many experiments on disordered materials, the vibrational density of states (VDOS) in the energy range $(2 \to 10 )\times 10^{-22} \; J$ exceeds significantly  beyond that of the phonon contribution. Referred as the boson peak, the functional  form of the excess VDOS is found to be universal for a wide range of amorphous systems irrespective of their microscopic details (i.e with different chemical bonding and short-range order structure, specifically for the close packed metals and the covalent networks) \cite{buch2,vdos}. The primary objective of this study is to seek this functional form using standard statistical tools.

 Previous attempts to understand the physical origin of the boson peak and other glass anomalies are usually based on various phenomenological models of the local interactions \cite{dpr, ell3} e.g. whether the underlying  structural disorder is of harmonic or anharmonic type. In the models based on harmonic degrees of freedom e.g. coupled harmonic oscillators with randomized force constants,  the excess VDOS  marks the transition between acoustic like excitations and a vibrational spectrum dominated by disorder \cite{sdg, tara, grig}. On the contrary, the soft potential model (SPM) suggests the vibrations of atoms in the strongly anharmonic potentials (the non-acoustic quasi local vibrations or QLV)  as the cause \cite{spm, gure, psg}.  Based on numerical simulations, the heterogeneous elasticity theory suggest the boson peak to originate from the spatial  fluctuations of the elastic constants on a microscopic length scale \cite{schi2}.  A first order transition theory relates the BP to the dynamics of domain walls that separate cooperatively rearranging regions \cite{lw}. Another theory predicts  the characteristic vibrations of nanometric clusters as the cause of boson peak; the vibrations are believed to exist e.g due to spatially inhomogeneous cohesion in glasses \cite{du, mns,sksq}. Analytical attempts to derive VDOS, based on a random matrix modelling of the dynamical (or Hessian) matrix of an amorphous solid, have also been considered e.g. \cite{grig, ml, smmm, bp,zac}; for example, the harmonic random matrix models (HRM) of Hamiltonian or dynamical matrix suggest the boson peak to originate from purely harmonic elastic disorder \cite{sdg, schi2, grig, srs, gual, zac}. The excess VDOS is also suggested as the disorder induced modified form  of Van Hove singularities of the  crystal \cite{chum}.  Research during last two decades  has introduced some new approaches too e.g. models  based on effective medium  theories \cite{wyt, degi} of marginal elastic instability  near a characteristic frequency or on  randomly jammed particles at zero temeperature interacting through pairwise potentials\cite{osln, mizu}.

 Although leading to many important physical insights in the low temperature anomalies,  a weak aspect of almost all these models is a lack of unanimity and their  dependence on widely different assumptions of the underlying interactions and nature of disorder. The need to establish the veracity of these assumptions motivated large scale numerical simulations which not only indicated failure of  many quantitative predictions of these models but fail to reach to a consensus too. For example, for very low frequencies $\omega < \omega_0$,  recent SPM models theoretically predict the VDOS behavior as $\rho(\omega) \propto {\omega^4 \over \omega_{bp}^3}$ \cite{gure, psg}; here $\omega_0$ is a characteristic frequency of the solid. While agreeing with qualitative prediction of the SPM model, the numerical study \cite{mizu}, based on the model of pairwise interacting randomly jammed particles, reports quantitative differences. On the contrary, the effective medium theories   (EMT) based on marginal instability near boson peak differ from both SPM model as well as \cite{mizu}  and  predict the low frequency 
  VDOS behavior as $\rho(\omega) \propto {\omega^2 \over (\omega^*)^{2}}$ (for $\omega < \omega_0$), significantly  bigger than the Debye prediction $\rho(\omega) ={\omega^2 \over (\omega^*)^{3/2}}$ \cite{wyt, degi} with $\omega^*$ as another characteristics frequency of the material. A similar non-Debye scaling of VDOS  is also reported by the replica theory approach \cite{fpuz}.  Although the  experimental study \cite{lern} indicates $\omega^4$ dependnece of VDOS at low $\omega$, it does not rule it out  as an artefact of the cooling process.   The numerical study in \cite{rpcv} claims a disagreement of most of the observations with results based on HRM models \cite{srs,gual} but agreement  with QLV model \cite{spm} without using adjustable parmeters.

The discord continues in high frequency regime too. EMT assuming marginal stability indicate a characteristic pleateau of VDOS at $\omega > \omega^*$ \cite{wy1, wy2, degi} 
A similar behavior is also indicated by the numerical analysis of a jammed particle model \cite{mizu}   although its result in low frequency regime is inconsistent with \cite{wyt, degi}. The existence of such a pleateau however is not reported in experimental analysis of amorphous materials e.g. \cite{ya26, ya27, ya21, ya28, yanno, hsgms} or in Euclidean random matrix approach of \cite{grig}; the latter study indeed suggest a semicircle form of VDOS with convex tails.  The experimental study of doped crystals indicates a Gaussian form of boson peak \cite{hsgms}

The exact role of the disorder in various anamolies also adds to further confusion. Although relevant for transport properties, strong experimental evidence indicates its relative insignificance in controlling the Boson peak anamoly.  Furthermore, even if justified for some systems, the applicability of the assumptions to wide range of amorphous solids and disordered lattices, where boson peak in observed, is not directly obvious or experimentally observed. Based on experimental evidence, it is expected that for solids with same local structure, irrespective of their long range order i.e crystalline or non-crystaline order,  should have similar boson peak amplitude.  Besides, it is not sufficient only to know why a particular feature appears in amorphous solids but also why it is absent in crystals. This motivates us to attempt in the present work a theoretical route based on the interactions, experimentally established  to be omnipresent at microscopic level and not hypothetical, of the smallest constituents of a generic solid i.e atoms and molecules and seek at to how the reorganization of the local units leads to emergence of collective interactions leading to different physical behavior in amorphous solids.

An inherent suggestion lurking behind many of the experimental observations is that 
the peak originates from the fundamental interactions which occur  in the energy range corresponding to vibrational molecular spectrum and thus give rise to  additional density of states (besides the phonon contribution) \cite{bb1,bb2,bb3}. 
The promising candidate in this context are  the molecular interactions  at medium range order (MRO) of glasses and their phonon induced coupling at larger length scales \cite{vdos, du, ell3, degi, sksq,mg, bb1, bb2, bb3}. As revealed by many studies in past, the interactions among the molecules separated by distances less then medium range order  are dominated by the Vanderwaal forces \cite{isra,ajs}. At larger distances however a collective interaction of the cluster of molecules can lead to emergence of new, modified form of forces \cite{bb3,arg,meek,lhe}. A conspiracy between the two types, acting at short and large length scales, can give rise to an arrangement of molecules as a collection of sub-structures e.g spherical clusters  in an amorphous solid of macroscopic size. The idea was introduced in a recent theory, describing an amorphous solid of macroscopic size as a collection of nano-size  sub-units  \cite{bb1,bb3,bb3}.  The latter, referred as basic blocks, are subjected to a phonon mediated coupling with each other through their stress fields (arising due to external perturbations); the coupling has an inverse cube  dependence on the distance between the block centers \cite{vl,dl,lg1}. The interactions among molecules within a single block  are however dominated by Vanderwaal forces of $1/r^6$ type (with $r$ as the distance between the molecules). 

Based on the basic block approach, the properties of  a macroscopic  amorphous solid can be described in terms of those of the basic blocks; for example, the approach was used in \cite{bb2} to explain the universality of the internal friction at low temperature. As discussed in \cite{bb1,bb2,bb3}, the properties of a basic block can be derived from  the molecular properties which are independent of system-specifics and thereby  lead to universalitiy in the low temperature properties e.g  specific heat, internal friction, coupling-strength ratio \cite{bb1,bb3,bb3}. The success of our approach at low temperature and nano length scales, renders it relevant to question whether it is applicable at high $T$ too or would it fail similar to TLS models? This is imperative therefore to analyze the viability of the theory to decipher  the boson peak orign and is the focus of present work.

As discussed in \cite{bb3}, a basic block is small in size (of the order of few inter-molecular distances)  and  contains approximately $8$ molecules,  independent of the system-specifics.  Although  in \cite{bb3}, the shape of the block was assumed spherical, it can be generalized to other shapes without any qualitative effect on its physics.  With a small number of molecules distributed within a radius of $\sim 2 \; nm$, the pairwise interaction strength bewteen them  can be asumed to be almost equal for all pairs; (the assumption is further supported by the many body aspect of intermolecular interactions). This in turn leads to representation of the block Hamiltonain by a  dense (or full) random matrix if the basis is chosen to be the {\it non-interacting basis} i.e product basis of the single molecule states. These considerations were used in \cite{bb2} to derive the ensemble averaged density of states $\langle \rho_1(e) \rangle$ for the basic block; the analysis therein indicated that  the level-density $\langle \rho(e) \rangle$  has a universal form of a semi-circle with its peak  at $x \approx 1/b$  and  the bulk of spectrum  lying  between  $0 \le e \le 2/b$. The parameter $b$ is determined from the molecular properties and is of the order of $10^{18} \; J^{-1}$ for a wide range of amorphous systems.

Neglecting the phonon mediated coupling among the blocks, the density of the states of the macroscopic solid can be  obtained by a convolution of those of the basic blocks. Using the standard central limit theorem, it is technically  straightforward to show that a convolution of the semi-circles  gives rise to a Gaussian distribution. Although this approach seems to justify the appearance of a peak in the density of states of the macroscopic solid, it is based on the assumption of  independent blocks.  The inclusion  of block-block interaction however invalidates the convolution route and a knowledge of block DOS is no longer sufficient. The DOS of the superblock can however be derived by   the standard route i.e by expanding  the Green's function in terms of the trace of the moments of its Hamiltonian. The latter is discussed in detail in \cite{vl,dl, bb2,bb3} and is briefly reviewed in section II. The derivation of its matrix elements, in the non-interacting many body basis of the basic blocks, and  their statistical behavior is discussed in detail in section III and IV, respectively.  The information is used in section V to derive the average DOS. The next section presents a comparison with experimental results. We conclude in section VII with a summary and discussion of our results.

\section{Superblock Hamiltonian}
We consider a macroscopic sample of an amorphous solid, referred here as the superblock. Following the standard route, the Hamiltonian $H$ of a solid of volume $\Omega$ can in general be written as the sum over intra-molecular interactions as well as inter-molecular ones
\begin{eqnarray}
H= \sum_k h_k({\bf r}_k) + {1 \over 2} \sum_{k,l}  {\mathcal U}(|{\bf r}_k - {\bf r}_l |)
\label{hat0}
\end{eqnarray}
with $h_k$ as the Hamiltonian of the $k^{th}$ molecule  at position ${\bf r}_k$
and ${\mathcal U}$ as a pairwise molecular interaction with arbitrary range $r_0$.

The superblock Hamiltonian can however be represented by an altrenative form.
As discussed in \cite{bb3}, the experimentally observed medium range ordering in amorphous systems permits $H$ to be described  as a collection of basic blocks too. The inter-molecular interactions can now be divided into two types (i)  among molecules within a block, referred as the ''self-interactions'' or intra-block ones, and, (ii) from one block to another, referred as the ''other body'' type or inter-block type. 
The Hamiltonain$H$ of the superblock can then be expressed as  a sum over basic blocks of volume, say $\Omega_b$,
\begin{eqnarray}
H= \sum_{s=1}^g {\mathcal H}^{(s)} + {1 \over 2} \sum_{s,t=1}^g \sum_{k \in s, l \in t}  {\mathcal U}(|{\bf r}_k - {\bf r}_l |)
\label{hat1}
\end{eqnarray}
where  ${\mathcal H}^{(s)}$ is the Hamiltonian  of a basic block labeled $''s''$, basically sum over the  ''self-interactions'' i.e molecular interactions within a block.
Here $g$ is the total number of blocks,  given by $g={\Omega \over \Omega_b}$ with $\Omega_b$ as the volume of the basic block. As discussed in \cite{bb1, bb2, bb3} and also mentioned in section I, a basic block size is typically of medium range order of the material, with $\Omega_b \sim 10^{-20} \; cm^3$. Consequently a superblock of typical experimental size $1 \; cm^3$ consists of $g \approx 10^{20}$ basic blocks.

As discussed in \cite{arg,meek,lhe,bb3}, the net molecular force of one block on another can also be described by an effective stress field.  Assuming the isotropy and the small block-size, it can be replaced by an average stress field at the  center, say ${\bf R_s}$, of the block.  In presence of an elastic strain field, say $e_{\alpha \beta}$ e.g of the phonons in the material,  the stress fields of the blocks interact with phonons strain field and $H$ in eq.(\ref{hat1}) can then be expressed as the sum over phonons contribution, say $H_{ph}$,  sum over those of non-interacting blocks and the terms describing stress-strain interaction (i.e coupling between phonons and blocks). Let $\Gamma^{(s)}_{\gamma \delta} ({\bf r})$ be the stress tensor at point ${\bf R_s}$ of the basic block ''s'', $H$ can then be written as 

\begin{eqnarray}
H= H_{ph} +  \sum_{s=1}^g {\mathcal H}^{(s)}  + \sum_{s=1}^g e_{\alpha \beta} \; \Gamma^{(s)}_{\gamma \delta}
\label{hatt1}
\end{eqnarray}
Note the last term in the above equation takes into account the molecular-interactions between blocks in eq.(\ref{hat1}).

As the strain tensor $e_{\alpha \beta}$ contains a contribution from the phonon field, the exchange of virtual phonons will give rise to an effective (RKKY-type)  coupling between the stress tensors of any two block-pairs. 
The total phonon mediated coupling among all blocks can then be  approximated as  \cite{vl, dl}
\begin{eqnarray}
V  =\sum_{s,t; s\not=t} V_{st} &=&  {1\over 8 \pi \rho_m v_a^2} \; \sum_{s,t; s\not=t} \; 
\sum_{\alpha \beta \gamma \delta} \;  
{ \kappa^{(st)}_{\alpha \beta \gamma \delta} \over   | \; {\bf R_s}-{\bf R_{t}} \; |^{3} }  \; \;   \Gamma^{(s)}_{\alpha \beta} \otimes
\;  \Gamma^{(t)}_{\gamma \delta}  
\label{zq++}
\end{eqnarray}

with  $\sum_{s,t}$ as the sum over all basic blocks, $\rho_m$ as the mass-density and $v_a$, ($a=l,t$), as the speed of sound in the amorphous material in  logitudinal or transverse directions. Here  the subscripts $\alpha \beta \gamma \delta$ refer to the tensor components of the stress operator, with notation $\sum_{te}$ implying a sum over all tensor components: 
$\sum_{te} \equiv \sum_{\alpha \beta \gamma \delta}$. 
The directional dependence of the interaction is represented by $\kappa^{(st)}_{\alpha \beta \gamma \delta}=\kappa^{(st)}(\theta, \phi)$; it is assumed to depend only on the relative orientation ($\theta, \phi$) of the block-pairs and is  independent from their relative separation \cite{dl}:

\begin{eqnarray}
\kappa^{(st)}_{ijkl} &=& \nu_0 \sum_{ijkl} \left(\delta_{jl} \delta_{ik} +  \delta_{jk} \delta_{il}  \right)  + \nu_2 \; \delta_{ij} \delta_{kl} 
 - 3 \; \nu_1   \sum_{ijkl} \left(n_j n_l \delta_{ik} + n_j n_k \delta_{il} + n_i n_k \delta_{jl} + n_i n_l \delta_{jk} \right)  \nonumber \\
&- & 3 \; \nu_2 \; \sum_{ijkl} \left( n_i n_j \delta_{kl} + n_k n_l \delta_{ij} \right)  
+ 15 \; \nu_2 \; \sum_{ijkl} \; n_i n_j n_k n_l  
\label{ad3}
\end{eqnarray}

where $\nu_0 =- {1\over 2} \; {v_t^2 \over v_l^2}$. $\nu_1=-{1\over 4}+ \nu_2$, $\nu_2= {1\over 2} \left(1-{v_t^2 \over v_l^2}\right)$ and ${\bf n} = n_1 \hat i + n_2 \hat j + n_k \hat k$ is the unit vector along the direction of position vector ${\bf r-r'}$.

As clear from the above, due to emerging interactions of the stress fields of block-pairs in presence of phonons, eq.(\ref{hatt1}) can again be rewritten as $H = H_{ph} + H_{nph}$ \cite{bb2,vl} with 
\begin{eqnarray}
H_{nph} =H_0 + V.
\label{zq}
\end{eqnarray}
Here $H_{nph}$ describes the non-phononic contribution to $H$, with $V$ as the net pair-wise interaction among the basic blocks given by eq.(\ref{zq++}) and $H_0$ as the total Hamiltonian of $g$ uncoupled basic blocks  
\begin{eqnarray}
H_0 &=& \sum_{s=1}^g \;   {\mathcal H}^{(s)}    
\label{zq+} 
\end{eqnarray}
with ${\mathcal H}^{(s)}$ same as in eq.(\ref{hat1}). With our interest in this work  in derivation of the vibrational density of states (VDOS) exceeding that of phonon contribution, we henceforth focus on $H_{nph}$ part of $H$ only.

\section{Superblock matrix elements} 

Our next step is to consider the matrix representation of $H_{nph}$; for notational simplification $H_{nph}$ is henceforth referred as $H$.  The physically-relevant basis here is the product basis of the non-interacting (NI) basic blocks,  consisting of a direct-product of the single-block states  in which each block is  assigned to a definite single-block state. As each block can be in infinite number of energy states, this leads to an  infinite number of of product states. But due to a minimum energy cutoff on the vibrational dynamics, only few of these states are relevant and it is sufficient to consider a tuncated basis of size $N$.  The basis, later referred as the NI basis, has selection rules associated with a 2-body interaction; only two blocks at the most can be transferred by $H$ to different single-block states.
As a consequence, many matrix elements are zero and $H$ is a sparse matrix. 
This can be explained as follows. 

Let  $|k_s\rangle$ and $E_{k_s}$, with $k_s=1,2,...N$, be the eigenvectors and eigenvalues of an individual block, say "s":
\begin{eqnarray}
 H^{(s)}_0 \;   |k_s\rangle =E_{k_s} \;  |k_s\rangle.
 \label{hs1}
 \end{eqnarray}
This basis therefore  consists of $M=N^{g}$ vectors given by 
\begin{eqnarray}
|k \rangle  = \; \prod_{s=1}^g \; |{k_s}\rangle.
\label{hb}
\end{eqnarray}
Here $|k\rangle$ refers to a particular combination of $g$ eigenvectors (one from  each of the $g$ blocks), with $|{k_s}\rangle$ as the particular eigenvector of the $s^{\rm th}$ block which occurs in the combination $|k\rangle$.  

From eq.(\ref{zq}), the matrix elements of $H$ are sum of those of $H_0$ and $V$. With NI basis consisting of the product of eigenfunctions of basic blocks, $H_0$ is diagonal in this basis

\begin{eqnarray}
H_{0; kl}  = \langle k | H_0 | l \rangle = 
 \sum_{s=1}^g \; E_{k_s} \; \delta_{kl}
\label{h0klq}
\end{eqnarray}

With NI basis as the eigenfunction basis of $H_0$, the role of $V$ is to mix these energy-levels. From eq.(\ref{zq++}),  the matrix elements of $V$ in NI basis can be written as

\begin{eqnarray}
V_{kl} = \langle k | V | l \rangle  
=  {1\over 2} \; \sum_{s, t; s\not=t} \; \sum_{\alpha \beta \gamma \delta}    \; D ^{(st)}_{kl} \;   \; U^{(st)}_{\alpha \beta \gamma \delta} \;  \; \Gamma^{(s)}_{\alpha \beta; kl}  \; \; \Gamma^{(t)}_{\gamma \delta; kl}
\label{vklq}
\end{eqnarray}
where
\begin{eqnarray}
\Gamma^{(s)}_{\alpha \beta; kl} &=& \langle k | \; \Gamma^{(s)}_{\alpha \beta} \; | l  \rangle = \langle k_s | \; \Gamma^{(s)}_{\alpha \beta} \; | l_s  \rangle,  \label{gm1} \\ 
U^{(st)}_{\alpha \beta \gamma \delta} &=& {1\over 2 \pi \rho_m v_a^2} \; { \kappa^{(st)}_{\alpha \beta \gamma \delta} \over   | \; {\bf R_s}-{\bf R_{t}} \; |^{3} }  
\label{u}
\end{eqnarray}
with 
\begin{eqnarray}
D^{(st)}_{kl} =  \prod_{q=1; \not=s, t}^g \; \; \langle {k_q} | {l_q}  \rangle 
= \prod_{q=1; \not=s, t}^g \; \delta_{k_q, l_q}.
\label{dkl}
\end{eqnarray} 
Here  the $2^{nd}$ equality  in the above equation follows due to orthonormal nature of the eigenfunctions of a basic block.  

In general, the eigenfunction contribution from a block  to an arbitrary  basis state $|k \rangle$ in the NI basis can be same as those of others or  different.
For example, consider following four many body states: 
\begin{eqnarray}
|a \rangle = | k_1, k_2, k_3 \ldots k_{g} \rangle, \hspace{0.2in} |b \rangle = | l_1, l_2, k_3 \ldots k_{g} \rangle \nonumber \\ 
|c \rangle = | k_1, k_2, m_3 \ldots m_g \rangle, \hspace{0.2in} |d \rangle = | l_1, l_2, m_3 \ldots m_g \rangle.
\label{bs}
\end{eqnarray} 
Here the symbols $k_n, l_n, m_n$ refer to different states of the $n^{th}$ block. As clear from the above, states $|a \rangle, |b \rangle$ are different only in $2$-body space (i.e  basis consisting of the products of the eigenfunctions of ${\mathcal H}_1$ and ${\mathcal H}_2$), the states $|a\rangle$ and $|c\rangle$ are different in $(g-2)$-body space (i.e product basis of the eigenfunctions of ${\mathcal H}_3 \ldots {\mathcal H}_g$) but same in $2$-body space consisting of product of the eigenfunctions of ${\mathcal H}_1, {\mathcal H}_2$, the states $|a \rangle$ and $|d \rangle$ differ  in $g$-body space.  

As the potential $V$ contains the terms only of type $\Gamma^{(s)}_{\alpha \beta}  \; \Gamma^{(t)}_{\gamma \delta}$, a matrix 
element $V_{kl}$ is non-zero only if the   basis-pair $|k\rangle, |l \rangle $ have same contributions from at least $g-2$ or more blocks. 
Let us refer a  basis pair $|k\rangle, |l\rangle $, different in the eigenfunction contributions from $n$ basic-blocks, as an $n$-plet with $0 \le n \le g$; for example, in eq.(\ref{bs}), $|a \rangle, |b \rangle$ form a $2$-plet,   $|a\rangle$ and $|c\rangle$  form a $(g-2)$-plet and $|a\rangle$ and $|d\rangle$ form a $g$-plet.
For later reference, it is worth noting that a $k,l$-pair forming an $n$-plet corresponds to same contributions only from $g-n$  blocks.

Eq.(\ref{vklq}) can then be expressed in a more detailed form 

\begin{eqnarray}
V_{kl} 
& =& {1\over 2}   \sum_{s,t; s\not=t} \; \sum_{\alpha \beta \gamma \delta}     \;  \; U^{(st)}_{\alpha \beta \gamma \delta} \; \; \Gamma^{(s)}_{\alpha \beta; k_s k_s}  \; \; \Gamma^{(t)}_{\gamma \delta; k_t k_t}  \hspace{0.5in}  {\bf 0-plet}, \; ( \; k_s =l_s, \; \forall \; s=1 \to g),  \nonumber \\
&=& {1\over 2}  \sum_{t; t\not=s}  \sum_{\alpha \beta \gamma \delta}    \; U^{(st)}_{\alpha \beta \gamma \delta} \; \; \Gamma^{(s)}_{\alpha \beta; k_s l_s}  \; \; \Gamma^{(t)}_{\gamma \delta; k_t k_t} \hspace{0.6in} {\bf 1-plet},  \; (k_s \not= l_s, k_t = l_t ,\; \forall \; t \not=s),    \nonumber \\
&=& {1\over 2}  \sum_{s; s\not=t}  \sum_{\alpha \beta \gamma \delta}    \; U^{(st)}_{\alpha \beta \gamma \delta} \; \; \Gamma^{(s)}_{\alpha \beta; k_s l_s}  \; \; \Gamma^{(t)}_{\gamma \delta; k_t l_t}, \hspace{0.6in} {\bf 1-plet},\;  (k_t \not= l_t, k_s = l_s, \; \forall \; s \not=t),  \nonumber \\
&=&  {1\over 2}   \sum_{\alpha \beta \gamma \delta}    \; U^{(st)}_{\alpha \beta \gamma \delta} \; \; \Gamma^{(s)}_{\alpha \beta; k_s l_s}  \; \; \Gamma^{(t)}_{\gamma \delta; k_t l_t} \; \hspace{0.5in} {\bf 2-plet},\;  (  k_{s,t} \not= l_{s,t}, \;  k_r = l_r, \; \forall \; r \not=s, t),   \nonumber \\
&=& 0  \hspace{1.9in} \; {\bf n-plet, n>2} \;  (k_{r,s,t} \not= l_{r,s,t}, \; {\rm for \; any} \;r\not=s,t),
 \label{vkl1}
\end{eqnarray}

As clear from the definition of  an $n$-plet, the number of states  $|l \rangle$ forming a $1-plet$ with a given state $|k \rangle$ is $^{(N-1)}P_1 \; ^g C_1 \;$ (with notation $^a C_b \equiv {a!\over b!(a-b)!}$ and $^a P_b \equiv {a!\over (a-b)!}$);
 here  $^g C_1$ corresponds to the number of ways one of the $g$ blocks  can be chosen such that its contribution to state $|l \rangle$ differs from that of $|k \rangle$. But even that even that contribution can differ in  $(N-1)$ possible ways from that of the state $|k \rangle$.
 For example, if say $s^{th}$ block contributes differently to $|k\rangle ,|l\rangle$ i.e (i.e $|k_s \rangle \not= |l_s \rangle$) and if $|k_s \rangle =|1 \rangle$ then  $|l_s \rangle$ can be $2 \ldots N$ (as each block has $N$ possible states).  Extending the above argument, the number of $|l\rangle$-states forming $n$-plet with a fixed $|k \rangle$ can be given as $^{(N-1)}P_n \; ^g C_n \;$.

As clear from the above, although, for a given $k$, there are $N^g$ possible values of $l$ (as $V$ is a $N^g \times N^g$ matrix) but only $\left(1+^{(N-1)}P_1 \; ^g C_1 + ^{(N-1)}P_2 \; ^g C_2\right)$ of them lead to non-zero $V_{kl}$. As $k$ itself can take $N^g$ values, this gives the total number of non-zero matrix elements as $\approx g^2 \; N^{g+2}$. As a consequence, $V$ is a highly sparse matrix in the NI basis, with  $N^{g}(N^g-g^2 N^2)$ zero elements). 

From eq.(\ref{h0klq}), $H_0$ is a diagonal matrix in the NI basis. This, along with the sparsity of $V$, implies $H$ as  a sparse matrix with non-zero matrix elements coming from  $H$-assisted coupling of only those basis states which form $j$-plets with $0< j \le 2$ .   It is worth noting here that the above sparsity occurs  due to choice of the  basis consisting of the products of single block states.

An important aspect of the non-zero matrix elements of $V$ is that many of them are equal. This follows because  $V$ can transfer only two blocks to different single block states but the number of blocks $g$ available for pairwise interaction is much larger: $g >2$. More clearly, it is possible that $V_{kl} = V_{ij}$ even if $|k \rangle, |l\rangle, \not= |i \rangle, \; |j\rangle$. This can happen if 

(a) $|k\rangle$ and $|i\rangle$  form $2$-plets with $|l\rangle, |j\rangle$ respectively. The two blocks which contribute differently to  the $k,l$-pair also contribute differently to $i,j$ pair but the rest of  $(g-2)$ blocks which have same contributions to both $|k\rangle$ and $|l\rangle$,  they have same contribution to $|i\rangle$ and $|j\rangle$ too but different from those of $k,l$ pair (latter necesaray to keep $k, l, i, j$ as distinct basis states).

(b) $|k\rangle$ and $|i\rangle$  form $1$-plets with $|i\rangle$ and $|j\rangle$ respectively. One block which contributes differently to $k,l$-pair also contributes differently to $i,j$ pair. Here again the rest of $(g-1)$ blocks, with same contributions to both $|k\rangle$ and $|l\rangle$,  also have same contribution to $|i\rangle$ and $|j\rangle$  but different from those of the $k,l$ pair.  

For example, in  eq.(\ref{bs}),   $|a \rangle$ and $|c \rangle$ form $2$-plets with $|b \rangle$ and $|d \rangle$ respectively. For both $|a \rangle$ and $|b \rangle$, $g-2$ blocks are in same states $k_3,\ldots, k_g$. Similarly $|c \rangle$ and $|d \rangle$ have same contributions  from $g-2$ blocks but these are now in states  $m_3,\ldots, m_g$. Eq.(\ref{vkl1}) then implies  that 
\begin{eqnarray}
V_{ab} = V_{cd}
=  {1\over 2}   \sum_{te}    \; U^{(st)}_{\alpha \beta \gamma \delta} \; \; \Gamma^{(s)}_{\alpha \beta; k_1 l_1}  \; \; \Gamma^{(t)}_{\gamma \delta; k_2 l_2}. 
\label{vabcd}
\end{eqnarray}
Following from the above, the number of those $2$-plets which are equal can be $(N-1)^{g-2}$.  As a consequence, the number of binary correlations among matrix elements becomes very large and is almost of the same order as the total number of their pairs. As discussed later in section V, this information is relevant for the derivation of density of states of $H$.

Before proceeding further, it is natural to query about the appropriate size $N$ of the basis.
This is turn depends on the energy scales associated with the dynamics, thus making it necessary to first seek them.
 
\section{Relevant length and energy scales}

As discussed in \cite{bb3}, a change in the local structure due to dispersion interaction  of the molecules gives rise to  their strain fields. The existing phonon in the material mediate a coupling of the strain field of one molecule with the stress of another; this can also be expressed as the pairwise coupling of their stress fields.  As the latter is sustained by the energy of the dispersion intercations, this introduces a length scale $R_0$ \cite{bb3} where
\begin{eqnarray}
R_0^3 = {(1+y)^6 \;   \; A_H  \; M^2  \over    8 \;  \pi^2  \; N_{av}^2   \; \rho_m} \left({v_a \over  \gamma_a} \right)^2
\label{r03}
\end{eqnarray}
with $\rho_m$ and $v_a$ same as in eq.(\ref{zq++}), $A_H$ as a material-specific constant, referred as the Hamaker constant, $M$ as the molar mass, $N_{av}$ as the Avogrado's number and $y ={R_v \over R_m}\sim1$, with $R_m$ as the radius of a molecule. Further $\gamma_a$ is  the strength of longitudinal or transverse phonons induced $r^{-3}$ coupling of  the two molecules (latter referred by the subscript $m$). As discussed in \cite{bb3}, $R_0 \sim 3-4 \; \AA$ for typical amorphous material.

With $V$ describing an interaction of basic blocks and each such block a cluster of molecules, $R_0$ plays an important role in the dynamics generated by the Hamiltonian $H$ \cite{bb3}. For example, for  $R_b < R_0$ with $2 R_b$ as the linear size of a basic block,  the availability of  excess dispersion energy  results in scattering of phonons which adversely affect the coupling of blocks. And, for $R_b > R_0$, the phonon mediated coupling requires additional energy resources, the availablity of which at very low temperature is not obvious.This intuitively suggests $2 R_b$ as an optimum linear size of the basic blocks. As discussed in \cite{bb3}, the length scale $R_b$  can be expressed in term of the molecular parameters 
 \begin{eqnarray}
 R_b^2 =  { R_0^3\over 4 \; R_v}
 \label{ve0}
 \end{eqnarray}
 with $R_v$ as half of the average distance between two neighboring molecules. Typically $R_b \sim 10-15 \; \AA$ which is the length scale associated with medium range order in amorphous materials \cite{bb3} (specific values of $R_b$ for 18 glasses are listed in table I). 
 
 The above in turn introduces  an energy scale $e_{ir}$ (also referred as ''Ioffe-Regel energy''), in the dynamics at which phonon scattering takes place,
 \begin{eqnarray}
e_{ir} =  \hbar \omega_{ir}  \sim {\eta_{ir} \; \hbar \;  v_a \over 2 R_b}.
\label{eu}
\end{eqnarray}
with  $\eta_{ir}$ as a glass specific constant ($\sim o(1)$). Further, following from eq.(\ref{ve0}) and eq.({r03}), $R_b \propto v_a$ which renders $e_{ir}$  independent of the  longitudinal or transverse  aspect of sound velocity.  With typical sound velocity $\sim 10^3 \; m/s^3$ and $\hbar \sim 10^{-34} \; J-s$, the above gives $e_{ir} \sim 10^{-21} \; {J}$ (specific values of $R_b$ for 18 glasses are listed in table I).

The study \cite{jl} considers  the nature of phonon mediated coupling between two spins and indicates its dependence on their distance, say $R$, relative to the phonon wavelength $\lambda$. For $R < \lambda$, the copuling  varies as ${1/ R^3}$ but it develops an oscillatory behavior for  $R > \lambda$. This formulation was  later on generalized, in \cite{dl} to a phonon mediated coupling of the nanosize amorphous blocks. Following the ideas in \cite{dl}, the form of $H_{nph}$  given by eq.(\ref{zq}) with $V$ given by eq.(\ref{zq++}) is valid only for energy scales $e < e_{ir}$. For energy scales $e > e_{ir}$, the matrix elements $V_{mn}$ in NI basis become a rapidly oscillating function  of $m, n$.  The size $N$ therefore refers to total number of only those  eigenstates of $H_0$ whose energies are less than $e_{ir} $. The matrix elements $V_{mn}$ mixing two states $|m \rangle, |n \rangle$, with either of their energies higher than $e_{ir}$,  are therefore assumed to be neglected. The assumption restricts the applicability of the present analysis only to energies less than $e_{ir}$. 
Thus $e=e_{ir}$ marks the point on the spectrum where  non-phonon DOS (also referred as the excess DOS) is assumed to approach zero (as contribution from higher states assumed to be negligible).

Another energy scale present in  the low temperature behavior is related to phonon energy. Based on Debye theory, the oscillators coupled by phonons undergo collective vibrations for phonons energies $\hbar \omega < \hbar \omega_d$ and independent ones for $\hbar \omega > \hbar \omega_d$ with $\omega_d$ referred as the Debye frequency. The latter can be expressed in terms of the longitudinal and transverse phonon velocities in the material: ${1\over \omega_d^3} = { 1\over 18 \pi^2 \rho_n} \left({1\over v_l^3} + {2 \over v_t^3} \right)$ with $\rho_n$ as the number of oscillators per unit volume. 
With oscillators as the basic blocks in our analysis, we have  $\rho_n =1/\Omega_b$ with $\Omega_b=4 \pi R_b^3/3$  
and $\eta_d^3={27 \pi \over 2 \left(1+ 2\left({v_l\over v_t}\right)^3\right) }$. 
The Debye energy $e_d= \hbar \omega_d$ can now be expressed as,
\begin{eqnarray}
e_d = {\eta_d \; \hbar \; v_a\over R_b}  
\label{debye}
\end{eqnarray}
Based on $v_a$ and $R_b$ (taken from \cite{bb3}),  table I  lists the  $\eta_d$ and $e_d$-values  for 18 glasses  predicted by the above formulation along with corresponding experimental values for the known cases. As can be seen from the table, the  typical value of $e_d \sim 2 e_{ir} \sim 10^{-21} \; J$ .

A third energy scale affecting  the spectral statistics of the superblock Hamiltonain $H$ is  the  width of the spectrum of  a free basic block $e_{c} =\hbar \omega_{c}$.
As discussed in \cite{bb2},   $e_{c} = {2\sqrt{2}\over b}$ (given by eq.(25) of \cite{bb2}) where  
\begin{eqnarray}
b  \approx   \; { 36  \over   \eta  \; \sqrt{ z \; g_0} \; A_H}  \; {y^6 \over (1+y)^6}
=  { 9  \over 4 \; \sqrt{ 3 } \; A_H }  \; \left({y \over 1+y}\right)^{9/2}
\label{b1}
\end{eqnarray} 
with $y \sim 1$,  $z$ as the number of nearest neighbors of a given molecule, $g_0$ as the total number of molecules in the block ($g_0 \sim 8$) and $A_H$ same as in eq.(\ref{r03}). The $b$-values for $18$ glasses are listed in  table I of \cite{bb1}  which gives $e_{c} \sim 10^{-18} \; {\rm J}$.
 
With both $e_{ir}, {e_d}$ typically of the same order ($\sim 10^{-21} \; {\rm J}$), clearly $e_c$ far exceeds them. The former energies are however comparable to the edge bulk meeting  point $e_0$  of a basic block spectrum \cite{bb1}: 
\begin{eqnarray}
e_0 \approx {\sqrt{2}\over 3 \lambda b} \sim 10^{-21} \; {\rm J}
\label{e0}
\end{eqnarray}
 with $\lambda \approx 498$. This in turn suggest that,  below Debye energy,  the  phonon mediated coupling does not affect higher energy states of a  basic block and the contribution  to the DOS of the superblock comes only from the edge states or edge-bulk meeting region.

\begin{table}[ht!]
\caption{\label{tab:table I} {\bf Theoretically predicted parameters of excess VDOS:} ({\it Here all energy scales in ${\rm THz}$ units}). 
The table gives the theoretically predicted values for the variance $\sigma$ and mean $e_{bp}$, given by eq.(\ref{sig1}) and eq.(\ref{mu}) respectively, for the bulk DOS of a macroscopic sample for $18$ glasses. The values of speed of sound in the material  for each glass are taken from \cite{mb, pohl} and basic block size $R_b$ from \cite{bb3}. Here $M$ refers to the mass of  the basic structural unit participating in the dispersion intercations within a basic block of linear size $R_{b}$ \cite{bb1, bb2}.
For some cases, $R_{b}$ for longitudinal phonons can differ from that of  transverse ones \cite{bb3}.  This motivates us to display here  the values for $\sigma$ and $e_{bp}$  calculated for both length scales i.e for $L=2 R_{ba}$ in eq,(\ref{sig1}) , labelled as $\sigma_a, \mu_a$ with $a=l,t$ , respectively; here the subscript $a$ refers to the contribution from longitudinal or transverse phonons. 
 The last column displays the available experimental values for the boson peak location of some glasses, with  data-source given in square bracket (converted to ${\rm THz}$ with $1 \; {\rm cm^{-1}}=0.124 \; {\rm meV} \approx 0.03 \; {\rm THz}$ ). Note contrary to table I, here $R_b$ is determined from eq.(\ref{ve0}) with $\gamma_a$-values taken from \cite{mb}. We note the boson peak here refers to the peak in excess VDOS (instead of reduced VDOS $\rho(\omega)/\omega^2$). With mass $M$ appearing in the theoretical predictions, a deviation of the latter from the experiments can also be used to identify the correct basic structural unit dominating the dispersion interactions within a  basic block \cite{bb1,bb2}. }
  \begin{center}
\begin{ruledtabular}
\begin{tabular}{|l|c|ccccc|ccccccc|r|}
Index & Glass & $M\atop {gm \over mol}$ & ${\rho \atop \; {gm\over cm^3}}$ & ${R_{bl} \atop \AA} $ & ${v_l \atop  Km/s^3}$  &  ${v_t \atop Km/s^3}$ & ${e_0 \; \atop {Thz}}$ & $\eta$ & ${e_d \atop {Thz}}$ & ${e_{ir} \atop {Thz}}$  & ${\sigma \atop {Thz}}$ & ${e_{gh} \atop {Thz}}$ & ${e_{bp} \atop {Thz}}$ &  ${e_{bp,exp} \atop {Thz}} $    \\
\hline
 1 &  a-SiO2          &  120.09 &    2.20 &   10.03 &  5.80 &    3.80 & 1.58 &   1.74 &   10.03 &    5.02 &        0.84 &    4.77 & 2.51  &  1.51  \cite{stol}  \\
 2 &  BK7             &   92.81 &    2.51 &    9.72 &      6.20 &    3.80 &  1.85 &     1.64 &   10.44 &    5.22 &     0.87 &  4.96 & 2.61 & \\
 3 &  As2S3           &   32.10 &    3.20 &    9.27 &    2.70 &    1.46 & 4.50 &     1.46 &    4.25 &    2.12 &       0.35 &    2.02 & 1.06 & 0.75  \cite{nema}  \\
 4 &  LASF              &  167.95 &  5.79 &   10.32 &   5.64 &    3.60 &  3.79 &    1.70 &    9.27 &    4.64 &       0.77 &    4.40 &  2.32  & 2.32 \cite{prat}  \\
 5 &  SF4             &  136.17 &    4.78 &    9.35 &     3.78 &    2.24 &  2.10 &    1.59 &    6.42 &    3.21 &       0.53 &    3.05 & 1.60 & \\
 6 &  SF59            &   92.81 &    6.26 &    6.37 &     3.32 &    1.92 &  3.28 &    1.55 &    8.09 &    4.04 &       0.67 &    3.84  & 2.02  &  \\
 7 &  V52             &  167.21 &    4.80 &   10.04 &    4.15 &    2.25 & 2.09 &   1.46 &    6.04 &    3.02 &        0.50 &    2.87   & 1.51 & \\
 8 &  BALNA           &  167.21 &    4.28 &   11.36 &  4.30 &    2.30 & 1.72 &   1.44 &    5.47 &    2.73 &        0.46 &    2.60    & 1.37  &\\
 9 &  LAT             &  205.21 &    5.25 &   10.72 &    4.78 &    2.80 &  2.29 &  1.57 &    7.00 &    3.50 &        0.58 &    3.32    & 1.75  & \\
10 &  a-Se            &   78.96 &    4.30 &   11.81 &    2.00 &    1.05 &  3.34 &  1.42 &    2.40 &    1.20 &        0.20 &    1.14  & 0.60 & 0.52 \cite{novi}  \\
11 &  Se75Ge25        &   77.38 &4.35 &    NaN &     0.00 &    1.24 &   6.23 & 3.49 &     - &     - &     - &     - &     -& \\
12 &  Se60Ge40        &   76.43 & 4.25 &   8.82 &    2.40 &    1.44 &   6.80 & 1.60 &    4.37 &    2.18 &         0.36 &    2.08 & 1.09  &\\
13 &  LiCl:7H2O       &  131.32 &  1.20 & 13.28 &   4.00 &    2.00 &   1.19 & 1.36 &    4.08 &    2.04 &         0.34 &    1.94   & 1.02 & 1.74 \cite{kgb} \\
14 &  Zn-Glass    &  103.41 &    4.24 &    9.28 &    4.60 &     2.30 &   1.93 & 1.36 &    6.73 &    3.36 &         0.56 &    3.19  & 1.68  &\\
15 &  PMMA            &  102.78 & 1.18 &  15.45 &    3.15 &    1.57 &    1.53 & 1.35 &    2.76 &    1.38 &         0.23 &    1.31  & 0.69  & 0.44-0.58 \cite{ryz}\\
16 &  PS              &   27.00 &    1.05 &    9.09 &    2.80 &    1.50 &        1.51 & 1.45 &    4.46 &    2.23 &     0.37   & 2.12& 1.11  &   0.49   \cite{pala}    \\
17 &  PC              &   77.10 &    1.20 &   15.75 &    2.97 &    1.37 &      1.51 &  1.26 &    2.37 &    1.19 &   0.20 & 1.13 & 0.59  &\\
18 &  a               &   77.10 &    1.20 &   12.47 &    3.25 &    - &        1.50 & - &    - &    - &       - &     -  &  & \\
  \end{tabular}
\end{ruledtabular}
\end{center}
\end{table}

\section{Statistical properties of  matrix elements}

As mentioned  in section I, the complexity of intermolecular interactions within a basic block manifest through radomization of its Hamiltonian matrix. As the superblock Hamiltonian $H$ is a tensor sum of the basic block Hamiltonians perturbed by phonon mediated coupling, this  randomizes the matrix elements $H_{kl}$ too which in turn result in the fluctuations of its physical properties. The latter often originating from quantum aspects, their influence is important at low temperatures and could be  the key-mechanism behind the observed deviations of the low temperature non-crystalline properties from those of crystals. It is therefore necessary to describe $H_{kl}$ statistically. The superblock properties are  then best described by a matrix ensemble of the superblock Hamiltonians $H$ (instead of just a single $H$). Here the $H$-matrix for each superblock is expressed in the product basis of its own basic blocks states in non-interacting limit.

The statistical behavior of $H$ depends on  its two components i.e $H_0$ and $V$. 
As $H_0$ in the NI basis is a diagonal matrix with its eigenvalues $E_k$ as the diagonals i.e $H_{0;kk}=E_k =\sum_{s=1}^g E_{k_s}$. Here the energies $E_{k_s}$ of the $''s''$  basic block, in non-interacting limit, are independent random variables  with a  mean $\langle E_{k_s} \rangle$ and a variance $ (\langle E_{k_s}^2\rangle - \langle E_{k_s} \rangle^2)$ with $\langle . \rangle$ implying the averaging over an ensemble of basic blocks \cite{bb1}. 
As the superblock consists of many such blocks (their number $g \sim 10^{20} $ in a typical superblock of volume $1 \; cm^3$), the central limit theorem  predicts $H_{0;kk}=E_k$ as a Gaussian variable with a mean $\mu_k$ and variance $\eta^2_k$ where 
\begin{eqnarray}
\mu_k &=& \langle E_{k} \rangle =\sum_{s=1}^g \langle E_{k_s} \rangle \approx g \; \langle E_{k_s} \rangle_{sp}, 
\label{avm0} \\
\eta^2_k &\approx&  \langle E_{k}^2\rangle - \langle E_{k} \rangle^2 = \sum_{s=1}^g \left(\langle E_{k_s}^2\rangle - \langle E_{k_s} \rangle^2\right).   
\label{avh0}
\end{eqnarray}
Here $\langle E_{k_s} \rangle$ refers to ensemble average of the specific energy state of the $s^{th}$ basic block (i.e single body state) that contributes to $k^{th}$ energy state of the super block. 
The $\sum_{s=1}^g $ then leads to a double averaging, over all basic blocks in a superblock as well as an ensemble of superblocks, indicated by  $\langle E_{k_s} \rangle_{sp}$.  

We note that $\langle E_{k_s} \rangle$ varies from bulk to edge by a factor of $10^{-3} \; J$ (with typical values of $E_{k_s}$ in the edge ($\le e_0$) and bulk as $\sim 10^{-21} \; J$ and  $\sim 10^{-18} \; J$ respectively). Due to airy function behavior,however,  a rare event in which $E_{k_s}$ take a large negative value in the edge is also probable. Further as $g$ is quite large, $E_k$ in the bulk can be very large   even in their non-interacting limit (maximum value of $E_k \sim g \; 10^{-18} \; J \sim 100 \; J$). 

As different many body energy states in general consist of different combinations of $g$  single body states and, as $g$ is quite large, $E_k$, even in the non-interacting limit,  increases rapidly from edge to bulk of the spectrum  ($E_k$ is  $\sim g \; 10^{-18} \; J \sim 100 \; J$ in the bulk  and $\sim e_0$ in the edge).  Consequently, $\mu_k \equiv \langle E_{k} \rangle$ is expected to vary with subscript $k$. For example, for an $E_k$ consisting of contributions  $ E_{k_s} \sim e_0$ for $s=1 \to g$ (i.e all contributing single block states in the edge-bulk meeting region, see \cite{bb1}), implies $\mu_k \approx g \; e_0 \sim 0.1 \; J$. But an $E_k$ with only  $E_{k_1}  \approx e_0$ and $ E_{k_s} \le 0$ for $s=2 \to g$ corresponds to $\mu_k \approx e_0 \sim 10^{-21} \; J$.

But, with $H$-spectrum confined to   $\le e_{ir}$, here we consider only the low lying many body states with total contribution from single block states $\le e_{ir}$. This in turn requires that each of  these  many body state  consists of single blocks states with almost all of them in edge or edge-bulk region ($\sim e_0$). We note here that although each single block spectrum has only $1-2$ states below $e_0$, the number of many body states with $E_k \sim e_0$ is very large; (for example, there can be $g$ many body states  which  correspond to $E_{k_1} \sim e_0, E_{k_s} \sim 0$, for $s=2 \to g$. As a consequence, the bulk of the many body spectrum is expected to lie in the region around $e_0$ with  mean $\mu_k \approx e_0 \sim 10^{-21} \; J$.

Our next step is to seek the distribution of the $V$-elements given by eq.(\ref{vklq}). This requires a prior knowledge about the statistical behavior of the stress matrix of each block.  As the latter  originates from the  dispersion type  intermolecular forces based on instantaneous dipoles \cite{bb3} in presence of the other blocks, it undergoes rapid fluctuations and gets randomized. This behavior is expected to manifest in its matrix representation in a physically motivating basis e.g. NI basis. The desired information can then be obtained by assuming the following for a basic block:

 (i) it is isotropic and rotationally invariant (as $L \gg a_0$ with $a_0$ as atomic dimension), 
 
 (ii) it is small enough to ensure homogenity of many body interactions within the block, leading to  $\Gamma^{(s) }_{\alpha \beta; mn} \equiv \langle m_s | \Gamma^{(s)}) |n_s \rangle \approx {\rm independent \; of} \;  {m_s, n_s}$  and therefore ${m,n}$, 
 
 (iii) the coupling between phonon and non-phonon degrees of freedom is a weak perturbation on the photon dynamics.

 (iv) the matrix representation of $\Gamma^{(s)}_{\alpha \beta; kl}$   in NI basis in a random matrix,  with almost all elements $\Gamma^{(s)}_{\alpha \beta; kl}$ behaving like  independent random variables, 
 
 (v) the stress fields of different blocks behave as independent random variables. 

Based on  the above assumptions, only following type of stress matrix elements correlations for a basic block  turn out to be non-zero \cite{bb2}:
\begin{eqnarray}
\langle \Gamma^{(s)}_{\alpha \alpha; mn}  \;  \Gamma^{(s) *}_{\gamma \gamma; mn}  \rangle &=& {N^{-1} \; {\omega}_{sb} \; \rho_m \; v_a^2 \; \Omega_b \; }  \; \langle Q^{-1}_a(\omega)  \rangle_{\omega} \; \delta_{\alpha \gamma}
\label{gcorr3}   \\
\langle \Gamma^{(s)}_{\alpha \beta; mn}  \;  \Gamma^{(s) *}_{\alpha \beta;  m n }  \rangle
&=& \langle \Gamma^{(s)}_{\alpha \beta; mn}  \;  \Gamma^{(s) *}_{\beta \alpha; m n}  \rangle= {N^{-1} \; {\omega}_{sb} \; \rho_m \; v_a^2 \; \Omega_b } \;  \langle Q^{-1}_a(\omega) \rangle_{\omega}  
\label{gcorr4}
\end{eqnarray} 
with $v_a$ as the speed of sound waves (with $a=l, t$ for longitudinal and transverse cases respectively) and $\rho$ as the mass-density of the material and $\Omega_b$ as the volume of the basic block. Here $\langle . \rangle_{\omega}$ implies an averaging over $\omega$:
 \begin{eqnarray}
 \langle Q^{-1}_a(\omega) \rangle_{\omega}  = {1\over \omega_{sb}} \int_0^{\omega_{sb}} Q^{-1}_a(\omega) \; {\rm d}\omega.
\label{avq}
\end{eqnarray}
with $\hbar \omega_{sb}$ as the upper cutoff of the energy  available to a sub-block Hamiltonian $H_0^{(s)}$. (We note here that in \cite{bb1} $\hbar \omega_{sb}$ was taken to be the full spectrum width $e_c$ (defined below eq.(\ref{b1})). But as mentioned below eq.(\ref{b1}) in section IV, now only part of the spectrum  below $e_0$ of each basic block contributes  to $H$ and thus it is appropriate to choose $\hbar \omega_{sb} \le e_{0} \sim e_{ir}$.

Further, as $t\not=s$ in eq.(\ref{vklq}) and correlation between stress-matrices of different block are negligible, we have 
\begin{eqnarray}
 \langle \Gamma^{(s)}_{\alpha \beta; kl}  \; \; \Gamma^{(t)}_{\gamma \delta; kl} \rangle = \langle  \Gamma^{(s)}_{\alpha \beta; kl} \rangle \; \; \langle \Gamma^{(t)}_{\gamma \delta; kl} \rangle = \tau_1^2 =0  \qquad t \not=s
\label{gcorr5}
\end{eqnarray}

The above statistical behavior of stress matrix elements in turn leads to  a Gaussian behavior for $V_{kl}$; this can be shown as follows. Referring the moments $\langle (V_{kl})^n \rangle$ as $L_n$ and substituting eq.(\ref{gcorr5})  along with eq.(\ref{u}) in the ensemble average of eq.(\ref{vklq}),  leads to  
\begin{eqnarray}
L_1 \equiv \langle V_{kl} \rangle
&=& {\tau_1^2 \over 4 \pi \rho v_a^2} \;  \sum_{s,t; s\not=t} \; {1\over |{\bf R_s}-{\bf R_t}|^3 }\; \sum_{te} \; D^{(s t)}_{kl} \; \; 
\langle  \kappa^{(s t)}_{te} \rangle  = 0
\label{vmean}
\end{eqnarray}

The $2^{nd}$ moment $L_2$ of $V_{kl}$ can similarly be derived. An ensemble averaging of the square of eq.(\ref{vklq}), and then applying eqs.(\ref{u}, \ref{gcorr3})) and eq.(\ref{dkl}) (latter gives $( D^{(s t)}_{kl} )^2  = D^{(s t)}_{kl}$) leads to 
(with details given in {\it appendix} A)

\begin{eqnarray}
L_2 = \langle \left( V_{kl} \right)^2 \rangle
&=&  N^{-2} \; \omega_{cs}^2  \; K^2 \;    \langle Q_0^{-1} ( \omega) \rangle_{\omega} ^2 \; \;   C_{kl} 
\label{vvar1}
\end{eqnarray}
with $K^2= \sum_{te, te'}  \; Y_{\alpha \beta \alpha' \beta'}  \;
Y_{\gamma \delta \gamma' \delta'} \;
\langle \; \kappa^{(st)}_{te}   \;  \kappa^{(st)}_{te'}  \; \rangle$
where $Y_{\alpha \beta \alpha' \beta'} = q \; \delta_{\alpha \beta} \delta_{\alpha' \beta'} +
\delta_{\alpha \alpha'} \delta_{\beta \beta'}  + \delta_{\alpha \beta'} \delta_{\beta \alpha'} $
and $q=1-{v_t^2\over v_l^2}$. Further  $C_{kl}$ is the square of the ratio of two volume type terms, just a dimensionless constant,
\begin{eqnarray} 
C_{kl}  &=& {1\over 4\pi^2} \; \; \sum_{s,t} \; {D^{(s t)}_{kl} \; 
\Omega_b^2 \over   | \; {\bf R_s}-{\bf R_{t}} \; |^{6} }. 
\label{ckl}
\end{eqnarray} 
As discussed in {\it appendix} B, 
$C_{kl} =0$ for $j >2$,  $C_{kl} = C_0,   {C_0 \over g},   { C_0 \over g^2}$  for $k,l$ pair forming a $j$-plet, with $ j=0,  1,  2$ respectively, and  
\begin{eqnarray}
C_0 &\equiv & \sum_{k,l =1}^{M}  C_{kl} = {1\over 4\pi^2} \;   \sum_{s, t;  s\not=t}  \; {\Omega_b^2 \over  | \; {\bf R_s}-{\bf R_{t}} \; |^{6}} \approx  g \; d_0.
\label{c0}
\end{eqnarray}
with $d_0$ a dimensionless constant: $d_0 \approx z/144$ with $z$ as the number of nearest neighbor blocks. The above gives the variance $\nu_{kl}$ of the matrix elements of $V$ as

\begin{eqnarray}
\nu_{kl} \; \; &\equiv  & \; \; \langle \left( V_{kl} \right)^2 \rangle_e - \langle V_{kl}  \rangle_e ^2  \nonumber \\
& = & 2 \; N^{-2} \; \omega_{cs}^2   \; K^2  \;  \langle Q_0^{-1} ( \omega) \rangle_{\omega} ^2 \; \; C_{kl} 
\label{nukl}
\end{eqnarray}
Thus the strength of $V_{kl}$ depends on $C_{kl}$ which in turn is governed by  the nature of the interactions among the basic blocks. Here $K$ is a constant, given by $K^2=(8/3)[-3+4s+16 q(q+s+q s -1)]$ \cite{vl}. The  typical values of  $q \equiv 1-{ c^2_t \over c^2_l} \simeq 0.6$ and $s \equiv { {\rm Im} \chi_l \over {\rm Im} \chi_t} \simeq 2.6$, give  $K^2=122.1$.

Proceeding similarly, the  higher order moments of $V_{kl}$, say $L_n = \langle (V_{kl})^n \rangle$ can be calculated. From eq.(\ref{vklq}), one can write
\begin{eqnarray}
L_n \equiv  \langle \left( V_{kl} \right)^n \rangle  = 
2^{-n} \; \sum_{s_1, t_1\atop s_1\not=t_1} \ldots \sum_{s_n, t_n \atop s_n\not=t_n} \; \sum_{te_1 \ldots te_n}  \langle \prod_{p=1}^n \left( D^{(s_p t_p)}_{kl}  \; U^{(s_p t_p)}_{\alpha_p \beta_p \gamma_p \delta_p}  \; \Gamma^{(s_p)}_{\alpha_p \beta_p; kl}  \; \Gamma^{(t_p)}_{\gamma_p \delta_p; kl} \right)\rangle 
\label{mvar}
\end{eqnarray}
As discussed in {\it appendix C}, the $L_n$ satisfies following relations,
\begin{eqnarray}
L_{2n} &\approx &  (2n-1)!! \; L_2  \label{l2n} \\
L_{2n+1} &\approx & 0 \label{l2np}
\end{eqnarray} 
The above implies that $V_{kl}$ is a Gaussian distributed matrix element with its mean and variance given by eq.(\ref{vmean}) and eq.(\ref{nukl}) respectively.

As mentioned in previous section, even if the matrix elements of $V$ are different in $g$-body product basis space, they can be same in $2$-body product basis space. This in turn results in non-zero binary correlations among them.
For example, for the states $|a \rangle, |b \rangle, |c \rangle, |d \rangle$ considered in previous section, one has
\begin{eqnarray}
\langle V_{ab} \; V_{cd}\rangle &=&\langle (V_{ab})^2 \rangle = \langle (V_{cd})^2 \rangle \label{veq1} \\
&=&  {1\over 4}   \sum_{te,te'}    \; U^{(12)}_{\alpha \beta \gamma \delta} \; \; U^{(12)}_{\alpha' \beta' \gamma' \delta'} \; \Gamma^{(1)}_{\alpha \beta; k_1 l_1} \; \Gamma^{(2)}_{\gamma \delta; k_2 l_2} \; \Gamma^{(1)}_{\alpha' \beta'; k_1 l_1} \; \Gamma^{(2)}_{\gamma' \delta'; k_2 l_2} 
\label{veq2}
\end{eqnarray}

Following from the above, the probability density  of the diagonal elements of $H$ is given by a convolution of the Gaussian distributed $V_{kk}$ and $H_{0;kk}$ and is again a Gaussian.  Further as $H_{kl}=V_{kl}$ for $k\not=l$, the  off-diagonals of $H$ are  again Gaussian distributed, with a non-zero variance $\nu_{kl}$ only for the pairs $k,l$ forming an $n$-plet (for $0 \le n \le  2$). The distribution of the superblock Hamiltonian can therefore be well-represented by a Gaussian ensemble of sparse matrices, with mean and variance given as follows. From eq.(\ref{zq}), $H_{kl} =H_{0;kl} + V_{kl}$ and therefore 
\begin{eqnarray}
\langle H_{kl} \rangle &=&\langle H_{0; kl} \rangle + \langle V_{kl} \rangle = \mu_k \; \delta_{kl}, \nonumber \\
\langle H_{kl}^2 \rangle - \langle H_{kl} \rangle^2 &=& \langle H_{0; kl}^2 \rangle -\langle H_{0; kl} \rangle^2 + \langle V_{kl}^2 \rangle = \eta^2_k \; \delta_{kl} + \nu_{kl}
\label{hdis}
\end{eqnarray}
where  $\mu_k$ and $\eta^2_k$ are defined in eqs.(\ref{avm0},\ref{avh0}).

 It is worth noting that the sparsity of $H$ is governed by the strength of  phonon mediated coupling. Further, as mentioned in previous section, the total number of non-zero off-diagonals is  $\approx g^2 \; N^{g+2}$ which is much bigger than the total number $N^g$ of the diagonals. The statistical properties of the ensemble of superblock Hamiltonians should therefore  be dominated by the off-diaogonals i.e by the stres-stress interaction between the blocks and not by their nature. This may explain the universal behavior of glasses in the low temperature regime.

\section{Average Density of States (DOS)}

The randomization of the matrix elements of $H$, discussed above, reflects on its eigenvalues too, resulting  in the fluctuations of its DOS. But $H$ being a many body Hamiltonian, the fluctuations are expected to have negligible effect on the low temperature physical properties and the knowledge of an ensemble averaged  DOS is sufficient. Using the standard formulation, the many body DOS $\rho(e)$  of the Hamiltonian $H$  (eq.(\ref{zq})) can be expressed as $\rho(e) =  \sum_{n=1}^M \delta(e-e_n) $ with $e_n$ as the many body levels, with $M=N^g$ as the size of matrix $H$. An averaging of $\rho(e)$,  at a fixed $e$, over all replicas of $H$ in the ensemble then leads to its ensemble average at $e$: $\langle \rho(e) \rangle = \sum_{n=1}^M \langle \delta(e-e_n) \rangle$.

As discussed in \cite{bb1}, the ensemble averaged bulk density of states  of a single basic block, referred here as $\langle \rho_{1}(e)\rangle$ to distinguish it from that of many blocks,  turns out to be  a semi-circle in the bulk of the spectrum and a super-exponetially dcaying function in its edge. But as $H$ represents an amorphous solid of macroscopic size consisting of many such basic blocks, the pair-wise coupling among the latter is expected to modify $\rho(e)$ at least for large energies.  A modified density of states at macroscopic level is also expected on the grounds of  sparse structure of $H$ and can directly be derived, for energy scales $< e_{ir}$ (eq.(\ref{eu})), from its statistical behavior discussed in previous section (e.g. following the route discussed in section III of \cite{bb1}). However, based on the energy range of interest, it is technically easier to use alternative routes. The steps can briefly be described  as follows.

\subsection{Lower spectral edge}

For very low energies, the phonon mediated coupling between the blocks is very weak as compared to intra-block interactions; this can be seen by a comparison of a typical diagonal element, say $H_{kk}$ ($\approx \mu_k+ \sqrt{\eta_k^2+\nu_{kk}}$), with a typical off-diagonal, say $H_{kl}$ ($= V_{kl} \sim \sqrt{\nu_{kl}} \approx \sqrt{\nu_{kk}/g}$ with $g \sim 10^{20}$ for a superblock of volume $1 \; {\rm cm}^3$). The block-block interaction energy $V$ in eq.(\ref{zq++}) can then be ignored, rendering  $H \approx H_0$. (Alternatively this can  be seen from the second order perturbation theory of energy levels which gives change in energy due to perturbation as $\delta e_n =V_{nn} + \sum_{m,\not=n}  {|V_{mn}|^2 \over e_m-e_n}$. The perturbation $V$ mixes the energy levels if $\delta e_n \sim \Delta$, with $\Delta$ as the local mean level spacing of the energy levels of $H_0$. 
For $\delta e_n < \Delta$, the energy levels $e_n$ of $H$ can  then be approximated as those of $H_0$ i.e the sum of $g$ non-interacting single block states).

In  the lower energy regime (referred as lower spectral edge $- \infty  < e < e_{gl}$ with $e_{gl}$ a material dependent energy scale), the many body DOS $\rho(e)$ can then be obtained by a convolution of the single block DOS.
Consider the many body DOS $\rho_{g}(e)$ of a solid, consisting of $g$ basic  blocks; it can be expressed as the  covolution of many body DOS of $(g-1)$ basic blocks and a single basic block VDOS
\begin{eqnarray}
\rho_{g}(e) &=& \int \rho_{g-1}(x) \; \rho_{1}(e-x) \; {\rm d}x\label{mbd}
\end{eqnarray}
As  the blocks are  almost mutually  independent in the edge region, an
ensemble averaging of the above equation for this energy range can be expressed as
\begin{eqnarray}
\langle \rho_{g, edge}(e) \rangle =\int \langle  \rho_{g-1, edge}(x) \rangle \; \langle \rho_{1,edge}(e-x) \rangle\; {\rm d}x
\label{mbd1}
\end{eqnarray}

As discussed in \cite{bb1}, the ensemble averaged  edge DOS $\langle \rho_{1,edge}(e) \rangle$ for a single block Hamiltonian  in the edge region can be given as
\begin{eqnarray}
\langle \rho_{1,edge}(e) \rangle = {N \; b \over \sqrt{\lambda}} \; \; f(\lambda \; b\;e) \hspace{0.1in} - \infty  < e < e_0. 
\label{rlt}
\end{eqnarray}
with $b, e_0, \lambda$ defined near end of section IV,  and
\begin{eqnarray}
f(x) & \approx & x \; {\rm Ai}^2(-x)  + ({\rm Ai}'(-x))^2 + {1 \over 2} \;  {\rm Ai}(-x) \; \int_{-\infty}^{-x} {\rm Ai}(y) \; {\rm d}y  \label{a1l}  
\end{eqnarray}
with ${\rm Ai}(y)$ as the Airy function of the first kind. 

In principle, a substitution of eq.(\ref{a1l}) in eq.(\ref{mbd1}) leads to the many body VDOS; this is however technically complicated.  Fortunately, using Airy function asymptotics,  eq.(\ref{a1l}) can further be simplified as \cite{bb5}
\begin{eqnarray}
f(x) \approx {1\over 2} \; {\rm Ai}(-x)   \hspace{1.0in} x < e_0.
\label{fx2}
\end{eqnarray}

  To derive $\langle \rho_{g, edge}(e) \rangle$ in the regime $e < e_{ir}$, we proceed as follows. A double differentiation of eq.(\ref{mbd1}) with respect to $e$ leads to 
\begin{eqnarray}
{{\rm d}^2 \langle \rho_{g, edge}\rangle \over {\rm d}e^2} + e \; \langle \rho_{g, edge}\rangle  = I_g(e).
\label{mbd2}
\end{eqnarray}
where $I_g(e) = \int x \; \rho_{g-1}(x) \; \rho_{1}(e-x) \; {\rm d}x$. Assuming that the maximum of product $\rho_{g-1}(x) \; \rho_{1}(e-x)$ occurs at $x=\alpha \; e$ with $\alpha$ as a constant,  it can be  approximated as (details given in \cite{bb5})
\begin{eqnarray}
I_g(e) &\approx & \mu_0  \; \int  \rho_{g-1}(x) \; \rho_{1}(e-x) \; {\rm d}x =\alpha \; e \; \langle \rho_{g,edge}(e) \rangle.
\label{mbd3}
\end{eqnarray}

 This leaves eq.(\ref{mbd2}) in the standard differential equation form for the Airy function and, as a consequnce,  we have \cite{bb5}
\begin{eqnarray}
\langle \rho_{g,edge}(e) \rangle \approx M \;  {\mathcal C}_0 \; {\rm Ai}(-{\mathcal B} \; e)) 
\label{fx3}
\end{eqnarray} 
with  constants ${\mathcal C}_0$ and ${\mathcal B}$ determined from the normalization condition for the full DOS (i.e including bulk): $\int \langle \rho_{g}(e) \rangle \; {\rm d}e =M$. 
For later reference, we note that the total number of levels in the edge region is
$M_{edge} =  \int_{-\infty}^{e_{gl}} \langle \rho_{g,edge}(e) \rangle \; {\rm d}e \approx {2 \; M \;  {\mathcal C}_0 \over 3}$

As clear from a comparison of the above with eq.(\ref{rlt}) with $f(x)$ given by eq.(\ref{fx2}), the form of the many body DOS of the amorphous solid consisting of $g$ blocks remains unaffected in the edge region. This indeed reflects the weakness of phonon mediated coupling of the blocks at energies $e < e_0$.

\subsection{Bulk of the spectrum}

With phonons of higher energy coming into existence, the phonon mediated coupling between blocks becomes strong enough to perturb the energy states of $H$ significantly. Consequently, the latter can no longer be  assumed to be just  the sum over single block states and the convolution approach used above  to derive $\langle \rho(e) \rangle$ is not , in principle, valid. This motivates us to consider an alternative route, based on the standard Green's function formulation of the DOS: 
$\rho(e) = -{1 \over \pi} \; {\rm Im} \; G(z)$ with $G(z)={\rm Tr} {1 \over H -z} $ as the Green's function with $z=e-i\epsilon$. 
The ensemble averaged density of states can then be written as  
\begin{eqnarray}
\langle \rho(e) \rangle=- {1 \over \pi} \; {\rm Im} \; \langle  \; G(z) \; \rangle.
\label{rhoe}
\end{eqnarray}
Using the moments $T_n \equiv \langle {\rm Tr}\; H^n \rangle = {\int e^n \; \langle \rho(e) \rangle \; {\rm d} e \over \int \langle \rho(e) \rangle \; {\rm d} e} = {1\over M} \int e^n \; \langle \rho(e) \rangle \; {\rm d} e$, $\langle G(z) \rangle$ can further be expressed:
\begin{eqnarray}
\langle G(z) \rangle &=& \langle {\rm Tr} {1 \over H - z} \rangle 
=-{1 \over z} \sum_{n=0}^{\infty} \; {T_n\over z^n} \;  
\label{g4}
\end{eqnarray} 
As clear from the above, $\langle G(z) \rangle$ is the Steiltjes transform of $\langle \rho(e)\rangle$ for large $z$ defined as $\langle G(z) \rangle = \int {\langle\rho(t)\rangle \; {\rm d}t \over z-t} $. 
An inverse Stieltjes-Perron transform then gives, for $\langle \rho(e) \rangle$ continuous in the entire integration range, as  
\begin{eqnarray}
\langle \rho(e) \rangle = \lim_{\varepsilon \to 0} {\langle G(e-i \varepsilon) \rangle- \langle G(e+i \varepsilon) \rangle\over 2 i\pi}
\label{srho}
\end{eqnarray} 

For technical simplification, we shift the origin of the spectrum to $\mu$, defined as 
\begin{eqnarray}
{1\over M} \sum_{k=1}^M \langle H_{kk}\rangle ={1\over M} \sum_{k=1}^M \mu_k=\mu.
\label{amu}
\end{eqnarray}
The shifted eigenvalues, defined as $e'_n=e_n-\mu$, correspond to the Hamiltonian $H'=H-\mu I$ with $I$ as the identity matrix. The density of states $\rho(e')$ can again be defined by eq.(\ref{rhoe}) with the  moments  now given as $T'_n \equiv \langle {\rm Tr}\; (H')^n \rangle$. 
It is easy to calculate the first three moments. Following from eq.(\ref{hdis}), $\langle H_{kl} \rangle = \mu_k \; \delta_{kl}$, $\langle (H_{kl})^2 \rangle -\langle H_{kl} \rangle^2 = \eta_k^2 \;\delta_{kl}+\nu_{kl}$  with $\nu_{kl}$ given by eq.(\ref{nukl}). The above in turn gives 
\begin{eqnarray}
\langle H'_{kl} \rangle &=& (\mu_k- \mu) \; \delta_{kl} \approx 0, \label{hp1}\\
\langle (H'_{kl})^2 \rangle &=& (\eta_k^2 + \mu_k^2 -\mu^2) \;\delta_{kl}+\nu_{kl} \approx  \eta^2 \;\delta_{kl}+\nu_{kl}.
\label{hp2}
\end{eqnarray} 
with $\eta^2= {1\over M}\sum_{k=1}^M (\eta_k^2 + \mu_k^2 -\mu^2)$.
This leads to
\begin{eqnarray}
T'_0 &=& M, \hspace{1in}
T'_1 = \sum_{k=1}^M \langle H'_{kk} \rangle= 0  \\
T'_2 &=&  \sum_{k,l=1}^M \langle (H'_{kl})^2 \rangle = {M \; \sigma^2} 
\label{t2}
\end{eqnarray}
where $\sigma^2$ is obtained by substituting eq.(\ref{nukl}) along with eq.(\ref{c0}) in eq.(\ref{hp2}),
\begin{eqnarray}
\sigma^2  \approx   \eta^2  + 2 \; g \; d_0 \; e_{sb}^2 \; K^2 \; \langle Q_0^{-1} ( \omega) \rangle_{\omega}^2
\label{sig1}
\end{eqnarray}
 Substitution of $d_0 \approx 6/144$, $K^2 \approx 122$, $\langle Q_0^{-1} ( \omega) \rangle_{\omega} \sim 5 \times 10^{-4}$ in the above further gives $\sigma^2 \approx \eta^2 +10^{-6} \;g \; e_{sb}^2$. With $\eta \sim e_{ir}$ and taking $e_{sb} \approx  e_0 \sim e_{ir}$, the second term is relatively negligible, leading to $\sigma \approx \eta  \sim e_{ir}$.

The higher order moments,  can be obtained by expanding in terms of the matrix elements,
\begin{eqnarray}
T'_n =\langle {\rm Tr}\;H'^n \rangle &=& \sum_{ a_1, a_2,....,a_n}  \langle H'_{ a_1 a_2} \; H'_{ a_2 a_3} \; .....H'_{ a_{n-1} a_n} H'_{ a_n a_1} \rangle
\label{tn}
 \end{eqnarray} 
with subscripts $a_1, a_2,....,a_n$ referring to the states  in the $g$-body product basis of size $M=N^g$ and $a_j =1 \to M$.
As clear from the above,  the trace operation ensures that the terms always have a cyclic appearance. Further evaluation of $T'_n$ depends on the correlations among $n$ matrix elements; the dominant contribution comes from those types of correlations which are not only relatively larger in magnitude but also lead to higher number of terms in the multiple summation in eq.(\ref{tn}). 

Although an exact calculation of $T'_n$ including all possible correlations among the matrix elements  is technically complicated, fortunately,  in  $N\rightarrow \infty$,  only pairwise correlations will be of consequence. This can be explained as follows. As mentioned in previous section, following  standard central limit theorem, $H_{kl}$ can be described as a Gaussian variable with mean $\mu$. This in turn implies  $H'_{kl}$ too as a Gaussian variable but with zero mean; its higher order moments $T'_n$ can then be evaluated by  applying Wick's probability theorem or Isserlis theorem \cite{issr}  which states that

{\it if $X_1, X_2,\ldots, X_n$ are $n$ zero-mean Gaussian variables,  the ensemble average of their product is a sum over contributions from all possible pairings of $\{1,2,\ldots,n\}$ i.e all distinct ways of partitioning $\{1,2,\ldots,n\}$ into pairs $\{i,j\}$. The contribution from one such partition, say $p \equiv \{p_1,p_2,\ldots,p_n\}$ is the product over the averages of pairs in it}.

Thus for $n$ even, 
\begin{eqnarray}
\langle X_1 X_2 \ldots X_n \rangle = \sum_p \langle X_{p1} X_{p2} \rangle \langle X_{p3} X_{p4} \rangle \ldots  \langle X_{p(n-1)} X_{pn} \rangle
\label{xp}
\end{eqnarray}
Clearly,  if $n$ is odd,  there does not exist any pairing of $\{1,2,\ldots,2n+1\}$ i.e one of the terms always remains unpaired. With $\langle X_k \rangle=0$, Isserlis' theorem then  implies that $\langle X_1 X_2 \ldots X_{2n+1} \rangle=0$.

 It is important to note however that the binary associations  or pairs relevant here are those formed in $2$-body space.  Except for the requirement of binary association, it is clear that each $H$ appearing in $\langle H^{n} \rangle$ act for the most part on different single body spaces for $g \gg n$ (as  the  number of such spaces is very large in this case). This implies e.g.  $H_{ k_1 k_2}$ can be  equal to any of the rest $n-1$ elements (as discussed above eq.(\ref{vabcd})). As the number of binary associations of $n=2m$ objects  is $(2m-1)!!$, there exist $(2m-1)!!$ pair partitions of $\{1,2,\ldots,2n\}$: this yields ${(2m)! \over 2^m \; m!}$ terms in the sum. For example, for $n=4,6,8$ random variables,the total number of such terms is $3, 15, 105$ respectively. As a result, $T'_{2n}$ in large $n$-limit can be approximated as 
\begin{eqnarray}
T'_{2n} &\approx & (2 n-1)!! \; (T'_2)^n
\label{t2n}
\end{eqnarray}

Further, with $\langle H'_{kl} \rangle  = 0$, the above  implies that all odd order moments vanish:
\begin{eqnarray}
T'_{2n+1} \rightarrow  \; \; 0 
\label{t2n1}
\end{eqnarray}

Substituting the above results in eq.(\ref{g4}), we get 

\begin{eqnarray}
\langle G(z) \rangle &\approx & - {1 \over z} \; \sum_{n=0}^{\infty} \; (2n-1)!! \left({T'_2 \over z^2}\right)^{n}
 \label{g5}
\end{eqnarray}
Substitution of $T'_2$ from eq.(\ref{t2}) above, followed by an inverse Stieltjes-Perron transform then gives 
\begin{eqnarray}
\langle \rho_{bulk}(e') \rangle  
&=&    {M  \over  \sqrt{2\pi \sigma^2} } \;  {\rm exp}\left(-{e'^2 \over 2 \sigma^2} \right)
\label{gau}
\end{eqnarray}
with $\sigma$ given by eq.(\ref{sig1}). The above result is derived for a shifted spectrum, with $e'_n = e_n-\mu$ with $\mu$ defined in eq.(\ref{amu}).  The latter on substitution in eq.(\ref{gau}), leads to
\begin{eqnarray}
\langle \rho_{bulk}(e) \rangle  
&=&    {M  \over  \sqrt{2\pi \sigma^2} } \;  {\rm exp}\left(-{(e-\mu)^2\over 2 \sigma^2} \right)
\label{rhoe1}
\end{eqnarray}

To estimate $\mu$-value,  we note that  $\rho_{bulk}(e_{ir}) \to 0$ and $\rho_{bulk}(\mu) \gg  \rho_{bulk}(e_{ir})$ with $e_{ir}$ as the upper energy cutoff on the  Hamiltonain $H$. Further from eq.(\ref{fx3}), $\rho(0) \approx 0$ and, as discussed below eq.(\ref{avh0}),  the bulk of the many body spectrum is expected to lie in the region around $e_0$ with  mean $\mu \approx \mu_k \approx e_0 \sim 10^{-21} \; J$. This intuitively suggests that the center of the bulk lies at $e_{ir}/2$ and motivates us to conjecture
\begin{eqnarray}
\mu \approx {e_{ir}\over 2} \approx {\eta_d\over 4 \eta_{ir}} \; e_{d}.
\label{mu}
\end{eqnarray}

Following from the Gaussian form of $\langle \rho(e)\rangle$ above, we note  $\rho_{bulk}$  at $e=\mu+3 \sigma$ is ${{\rm exp}(-4.5)\over \sqrt{2\pi \sigma^2}}$ i.e $\approx 0.01$ times  its peak value. Recalling that 
$e_{ir}$ is the upper energy cutoff available to Hamiltonain $H$ i.e for non-phononic states, this suggests $e_{ir} \approx \mu+ 3\sigma$ and further implies, from eq.(\ref{eu}) and eq.(\ref{debye}), 
\begin{eqnarray}
\sigma & & \approx  {\eta_d\over \eta_{ir}} \; {e_{d}  \over 12}.
\label{sig3}
\end{eqnarray}
As can be seen from table II, the estimates for $\sigma$ as well as the conjecture for $\mu$, are indeed quite good at least for six glasses.

It is clear from the above that the bulk of the spectrum depends on the single parameter $e_{ir}$. With latter dependent on the  average properties of the many body inter-molecular interactions, it is not expected to vary much from one system to another; this is also indicated by the $\sigma$-values for $18$ glasses displayed in table I.

Eq.(\ref{rhoe1}) is derived without assuming any specific system information and is therefore applicable for a typical amorphous system. With typical speed of sound in glass $\sim 10^3 \; m/sec$ and $R_b \sim 10-15 \; \AA$, eq.(\ref{eu}) gives $e_{ir}$ (the fulll width   of the non-phonon spectrum of $H$)  as $\sim 12 \; {\rm meV} \approx 3 \; {\rm THz}$ which  is consistent with the range of non-phononic DOS observed in the range from $10^o K$ to $40^o K$ for different materials and also with the low temperature energy range of vibrational DOS.

\subsection{Higher spectral range}

The contribution to many body vibrational DOS from non-phonon vibrations, derived above for an amorphous solid, is based on a theory of coupled blocks of size $2 R_b$ (eq.(\ref{ve0})) with their interaction energy  given by eq.(\ref{zq++}).  With  $R_b$ sufficiently larger than atomic dimensions, the form of $V$ remains valid at higher energies too but it is now necessary to consider the phononic  contribution to manybody DOS. Although higher energy phonons with wavelengths $\lambda < 2R_b$  are subjected to strong scattering  but it does not result in their localization and energy is transported by diffusion \cite{allen}.  While,  for energy range $e > e_{gh}$, $\rho(e)$ is decaying as a Gaussian, the phonon contribution to DOS, referred as $\langle \rho_{debye} (e) \rangle $ compensates for the decay.  In the low temperature limit, the latter can be written as $\langle \rho_{debye} (e) \rangle \approx {9 \; e^2 \over e_d^3}$ with $e_d$  defined in eq.(\ref{debye}).

For $e > e_0$, the total VDOS  $\langle \rho_{total}(e) \rangle  = \langle \rho(e) \rangle   + \langle \rho_{debye} (e) \rangle  $  can now be expressed as

\begin{eqnarray}
 {1\over M}\; \langle \rho_{total}(e) \rangle  
&=&   {1  \over  \sqrt{2\pi \sigma^2} } \;  {\rm exp}\left(-{(e-\mu)^2 \over 2 \sigma^2} \right) + {9\; e^2 \over e_d^3} 
\label{gau2}
\end{eqnarray}
As clear from the above, the mean  $\mu$ gives the location of the maximum in the non-phonon DOS i.e the boson peak; using the standard notation $e_{bp}$ for the boson peak location, hereafter $\mu$ will be referred as $e_{bp}$.

Further it can be shown that the slope of $\langle \rho_{total}(e) \rangle $ approaches zero, implying a constant DOS, for $e > e_{gh} =\mu+x$ where
\begin{eqnarray}
x^2 \approx -2 \sigma^2 \;  \log \left[{18 \sqrt{2\pi} \left({\eta_d\over 12 \eta_{ir}}\right)^3 }\right]  
\approx -2 \sigma^2 \;  \log\left({0.026 \; \eta_d^3 \over \eta_{ir}^3}\right)
\label{xgh}
\end{eqnarray}
The constant pleateau is expected to continue upto $e \sim e_{ir}$ and  beyond that point, the DOS is expected to increase as $e^2$ similar to Debye DOS.  Using eq.(\ref{sig3}) for $\sigma$, an approximate theoretical prediction for $e_{gh}$ for a few glasses is given in table I. This is consistent with the behvaior predicted in \cite{degi} which  also suggests an onset of constant density at $e_{gh} \approx e_d(0.25 + -)$. Although a similar behavior is  predicted in  \cite{mizu} too however  their $e_d$-value seems to be different from ours.

\section {Comparison with experimental data}

Eq.(\ref{fx3}), eq.(\ref{rhoe1}) and eq.(\ref{gau2}) describe the behavior of VDOS in three different energy ranges. While the VDOS for higher energies  is just a continuation of the bulk form, the form of the edge level density is different from that of bulk and it is  important to know how and where they connect to ensure a gapless spectrum. We note that the Airy function in eq.(\ref{fx3})  can not behave as an appropriate function for the  probability density for $e >0$ as it starts oscillating. The bulk however is centered around $e=e_{bp} >0$. For the edge to  join the bulk smoothly, the Airy function must remain positive; the latter can be achieved by shifting the variable $e$ in eq.(\ref{fx3})   to $e-e_{bp}$.

The full theoretical form for the vibrational  DOS can now be given as
\begin{eqnarray}
 {1\over M}\; \langle \rho_{total}(e) \rangle  
 &=&   {\mathcal C}_0 \; Ai(-{\mathcal B} (e-e_{bp})+ {\mathcal D}_b \; e^2  \hspace{1.in}  e \le e_{gl}, \label{gauu4} \\
 &=& {1  \over  \sqrt{2\pi \sigma_0^2} } \;  {\rm exp}\left(-{(e-e_{bp})^2 \over 2 \sigma^2} \right) + {\mathcal D}_b \; e^2  \hspace{0.6in}  e > e_{gl}.
\label{gau4}
\end{eqnarray}
Here, the Gaussian now valid only in the regime $e > e_{gl}$,  its normalization is changed from
${1/\sqrt{2\pi \sigma^2} } \to {1/ \sqrt{2\pi \sigma_0^2} }$. 
Further the last  term in the right side corresponds to Debye contribution with ${\mathcal D}_b= {9\over e_d^2}$.

Referring  the  edge-bulk connecting point as $e_{gl}$,  
it must satisfy the relation $ \langle \rho_{bulk}(e_{gl}) \rangle = \langle \rho_{edge-l}(e_{gl}) \rangle$ which leads to
\begin{eqnarray}
{\mathcal C}_0 \; {\rm Ai}(-{\mathcal B} \; (e_{gl}-e_{bp}))  =
{1 \over  \sqrt{2\pi \sigma_0^2} } \;  {\rm exp}\left(-{(e_{gl}-e_{bp})^2 \over 2 \sigma^2} \right)
\label{egl0}
\end{eqnarray}
As $e_{gl}$ appears both sides, the equation can only be solved self-consistently/ numerically; we find
\begin{eqnarray}
e_{gl} \sim e_{bp} -\sigma \sqrt{2} \label{egl}
\end{eqnarray}
The above in turn leads to a relation between the normalizations of bulk and edge densities: ${\mathcal C}_0 \;\sqrt{2\pi \sigma_0^2} \approx {1\over 2.72 \;  {\rm Ai}(\sqrt{2}  \; \sigma \; {\mathcal B} ) }$. We note however that the shift of  the variable $e$ in eq.(\ref{fx3}) to $e-e_{bp}$ is not unique, if it is shifted instead  to $e-\mu_0$ with $\mu_0 > e_{gl}$ but arbitrary otherwise, it just affect the relative normalizations of the edge and bulk parts.

The experimental studies usually present the data for reduced DOS $\langle\rho_{total}\rangle/e^2$. To obtain the $\langle\rho_{total}\rangle$ from the data, we use {\it plotdigitzer} software to digitally scan the figures in \cite{ya21, ya25, ya26, ya27, ya28}), read the values for $\langle\rho_{total}\rangle/e^2$ for many $e$'s and convert them to $\langle\rho_{total}\rangle$ as well as excess DOS $\langle\rho_{total} \rangle-\rho_{debye}$. 
The converted data  is displayed in figures 1-7 along with our theoretical prediction in eq.(\ref{gauu4}) and eq.(\ref{gau4}). The latter requires a prior information about various energy scales. While theoretical prediction for $e_{ir}$ and$e_d$ are given by eq.(\ref{eu}) and eq.(\ref{debye}),  we only have an intuitive  information about $\eta_{ir}$ i.e $\eta_{ir} \approx \eta_{d}$. Using the conjecture, we have
\begin{eqnarray}
e_d \approx 2 \; e_{ir}  \approx 4 \; e_{bp}. 
\label{debye1}
\end{eqnarray}
Eq.(\ref{debye}) gives the theoretical formulation for  $e_d$ which along with above equation gives  $e_{ir}$ and $e_{bp}$. Table I lists theoretical  predictions for $e_d, e_{ir}$ and $e_{bp}$ for six glasses; the required $v_l, v_t$ for their determination are taken from experiments \cite{pohl, ya21, ya25, ya26, ya27, ya28}.   Here, due to a lack of data for the phonon mediated coupling strengths $\gamma_a$ for these glasses,   $R_b$ can not be determined from eq.(\ref{ve0}); we  use, instead, the theoreticaly predicted relation $R_b=R_0=4 R_m$ with $R_m$ as the molecular radius. To avoid cluttering the table II, the values for the edge density parameters ${\mathcal C}_0$ and ${\mathcal B}$ are given with the fitted function in each figure caption. We note that both $e_{ir}$ and $e_{bp}$ are indeed just fractions of one single scale i.e $e_d$. Using these parameters in eqs.(\ref{gauu4}, \ref{gau4}), the resulting distributions are then displayed in figures 1-7 alongwith their experimental counterparts.

Table II also displays the experimental values for $e_d, e_{ir}$ and $e_{bp}$ for each of the six glass. Here the experimental value for $e_{ir}$ refers to  the non-zero energy where  the excess DOS  effectively vanishes and experimental $e_{bp}$ corresponds to the maximum of excess DOS; both values are obtained from the experimental DOS data (displayed in figures 1-7). The table also lists the  theoretical predictions for the variance $\sigma$ and energy scales $e_{gl}, e_{gh}$ (the upper and lower limits for Gaussian predicted DOS given by eq.(\ref{xgh}) and eq.(\ref{egl}) along with  corresponding  experimental values; Here the latter for $e_{gl}$ and $e_{gh}$ are taken as the energies where experimental curve deviates from the fitted Airy function and Gaussian, respectively. 
As indicated by table II,  the relation $\sigma \approx e_{d}/12$, $e_{gh} \approx 1.9 e_{bp}$ is well satisfied by all six glasses. This seems to be consistent with \cite{mizu} which predicts  $\omega_{bp} \approx 2 \omega^*$; (we note $\omega^*$  of \cite{mizu} corresponds to our $e_{ir}$).

As clear from the figure 	1-6, not only  the Gaussian form is  consistent with experimentally observed  excess DOS in the bulk of the spectrum, the behavior in the lower edge also agrees well with Airy function prediction. We also note a difference of fitted Airy parameters $c_e, {\mathcal B}, e_{gl}$ as well as gaussian parameters $\mu$ and $\sigma$ for total DOS (displayed in part (a)) from that of excess DOS (displayed in part (b)); this is expected due to inclusion of the Debye contribution to DOS  in part (a) and its absence in part (b). Indeed the fitted parameters in part (b) are closer to their theoretical predictions.

For higher energies, we  find that,  a good agreement for total DOS i.e eqs.(\ref{gauu4}, \ref{gau4}) with experiments can be achieved only if the phonon contribution i.e Debye DOS is taken into account. Note here $e_d=\hbar \omega_d$ in principle should be obtained from eq.(\ref{debye}) but this requires a prior knowledge of $v_l, v_t$ as well as $R_m$ which in turn depend on experimental conditions as well the material specific glass structure. As the latter is not available to us, we use  $v_l, v_t$ values from \cite{mb, pohl} and, as expected, find $e_d$  to be different from those mentioned in   \cite{ya21, ya25, ya26, ya27, ya28}). However the $e_d$-values given by  fits in parts (a) of figures 1-7  and those mentioned in   \cite{ya21, ya25, ya26, ya27, ya28}) are in good agreement for Pb, Se, Glycerine and SiO$_2$ but differ by a factor of 2 for B$_2$O$_3$.

Figure 7 displays the comparison of experimental result  for additional excitations in $SiO_2$ (different source than that depicted in figure 6) with  a Gaussian fit; (the experimental data, shown as points, is adapted from \cite{buch} by using {\it plotdigitzer} software). As figure 7 indicates,  $e_{ir} \approx 3 \;{\rm THz}$ and $\sigma \approx 0.5 \; {\rm THz}^{-1}$ and $e_{bp} \approx 1.4 \; {\rm THz}$ which is consistent with eq.(\ref{eu}) and eq.(\ref{mu}) (later gives $e_{bp}=1.5 \; {\rm THz}$). Further, as metioned in \cite{buch},  the maximum of excess vibrational DOS in vitreous silica is $0.013 \; {\rm THz}^{-1}$. Although eq.(\ref{rhoe1}) gives the maximum ${1\over \sqrt{2\pi \sigma^2}} \approx 0.74 \; {\rm THz}^{-1}$, this discrepency  seemingly arises due to different energy range used for the numerical normalization of DOS in \cite{buch}. 

Previous experimental studies have reported a relation between Boson peak frequency and the bulk modulus of the medium. To explore this dependence, we first note that  $\rho v_t^2 =G$ and $\rho\left(v_l^2-{4\over 3} v_t^2 \right)=K$ with $G$ as the Bulk modulus coefficient.  Substitution of these relations  in eq.(\ref{eu}) leads to
\begin{eqnarray}
e_{ir,t} = {\eta_{ir,t} \; \hbar \over 2 \; R_b}  \; \sqrt{G \over \rho}, \qquad e_{ir,l} = {\eta_{ir,l} \; \hbar \over 2 R_b}  \; \sqrt{(K+(4/3) G)\over \rho}
\label{debye2}
\end{eqnarray}
With $e_{bp} \approx e_{ir}/2$, the above eqution is consistent with the experimental observation  $e_{bp} \propto \sqrt{G}$. To check the relation computationally, we  compute $G$ as well as $K$ for the five glasses from  the $v_l, v_t$ values listed in table I and plot it with respect to $e_{bp}$ for each glass (the mean value of the fitted Gaussian in figures 1-7); the result displayed in figure 8 confirms the above relation.

The good agreement between theory and experiments encourages us to theoretically predict the $e_{bp}$ values for 18 other glasses; these are listed in table I. Here the required values for $v_a$  for each case are taken from \cite{pohl,mb} and $R_b$, given by  eq.(\ref{ve0}), is calculated in \cite{bb3}. The table I also lists the experimental data for $e_{bp}$ for a few cases which seems to agree well with our prediction for some of them. The deviation in other cases could well originate from lack of information about $v_l, v_t$ values used in the experiments.

\section{Discussion and conclusion}

As the VDOS of an amorphous solid has been extensively researched in past, it is relevant to compare our approach and results with previous ones. 
Contrary to previous theories often based on various assumptions about nature of disorder and local interactions, the many block Hamiltonian considered in this paper is based on well-known dispersion interactions of molecules within clusters of MRO length scales and a phonon mediated stress interaction among these clusters; here the randomization of Hamiltonian originates from the complicated many body interactions.   With molecular clusters described as blocks, our approach is closer in spirit with those suggested in \cite{du, sksq, ell3} however there are some important differences e.g (i) contrary to \cite{du}, the blocks described in our approach are just nanoscale partitions of amporphous solid, (ii) the  clusters in \cite{du} can be of varying size but the size of our blocks is fixed. We also note that the potential formulation given by eq.(\ref{zq++}) is valid only for the  phonon wavelengths larger than atomic scales.  For smaller length scales i.e higher frequencies, the phonon mediated coupling of the blocks changes its form from a inverse cube dependnce on the distance to oscillatory form \cite{dl}. 

As the  randomization of the Hamiltonian in our analysis is caused not by any structural disorder but rather  due to instantaneous induced dipole interactions of the molecules  (dispersion interaction), our boson peak prediction is independent of the nature of structural disorder; this is consistent with observations indicating secondary role of disorder in vibrational spectrum properties \cite{chum, degi}. A comparison with  experimental data for the DOS of six amorphous solids confirms the prediction and encourages us to predict the location of boson peak for $18$ other amorphous solids (given in table I). We also find that the copuling between the blocks has no significant effect on the  VDOS in the edge of the spectrum and it retains the same approximate Airy functional form as that of a single basic block. Beyond a certain energy scale (of the order of Ioffe-Regel frequency), however phonon DOS becomes significant, resulting in a small flat region for a small energy range and thereafter dominating the DOS.

Extensive investigations for the functional form of the DOS, based on various models, numerical analysis as well as experiments, indicate three characteristic frequency regimes. In low frequency regime $\omega < \omega_0$, with $\omega_0$ as a characteristic frequency of the material,  the theoretical predictions are at variance and can broadly be divided into two categories: 
(i) the models based on soft localized vibrations predict  the DOS $\rho(\omega)$  to change, with increase in frequency $\omega$, from $\omega^4$ (for $\omega < \omega_0$) to $\omega^2$ dependence,  (for $\omega > \omega_0$). 
(ii) the models  suggesting an $\omega^2$ behavior (e.g. those based on effective medium theories \cite{degi,mmb}). For example, the numemerical study of \cite{mizu} based on jammed particles indicates presence of both soft localized modes violating Debye-scaling as well as phonon modes following Debye scaling  below $\omega_0$. On the contrary, effective mediium theories (EMT) based on marginal instability in amorphous solids subject to compression  \cite{degi} differs from \cite{mizu}, indicating  $\rho(\omega) \propto C \; \omega^2$  in regime $\omega \ll \omega_0$, with coefficient $C \sim (\omega^*)^{2}$ with $\omega^* \sim \omega_{ir}$. Although the VDOS in Debye's theory is also given as $\rho(\omega) \approx C_d \; \omega^2$, the difference lies in the coefficients: $C_d \propto   (\omega^*)^{-3/2} < C$. (We note that these  powers of $\omega$  indicate only local frequency-dependence; the exact functional form of $\rho(\omega)$ is more complicated). Although our analysis indicates an Airy function behavior for the low frequency regime,  a Taylor series  expansion of $Ai(-{\mathcal B} (e-\mu))$ near $e=\mu$ however reveals a power law dependence on $e$, with dominant power varying with distance $|e-\mu|$. We note that the limit $\omega_0$ varies from one model to another e.g $\omega_0 \approx 0.066 \; \omega_{bp}$ in \cite{mizu} and $\sim {\omega_{bp}^2\over \omega^*}$ in \cite{degi} with $\omega^*$ is another charateristic frequency scale  \cite{degi}; taking $\omega^* \sim \omega_{ir}$, this leads to  $\omega_0 \approx \omega_{bp}/2$. As discussed in section III, this is same as in our case with $\omega_0 \approx e_{gl} \approx \omega_{bp}/2$.

For intermediate frequency range $\omega_0 < \omega < \omega^*$, almost all models e.g \cite{srs}, those based on jammed particles \cite{mizu} as well as effective medium theories \cite{degi, wyt}   predict a local $\omega^2$ dependence of DOS. Some earlier studies however have also suggested a $\omega^{3/2}$ behavior \cite{grig}. Our prediction of a Gaussian form  for nonphononic DOS i.e excess DOS is consistent with a local behavior $\sim \omega^2$ near $\omega_{bp}$; the display in figure 1-7 confirm the consistency of the Gaussian form with experimental data too. 

In higher frequency range i.e $\omega > \omega^*$ with $\omega^*$ as another charactreistic frequency, the jammed particle models as well as effective mediium theories predict  a constant behavior  of the DOS; the latter however seems to be an artefact of the models.
As discussed in previous section, our analysis also indicates a constant DOS where it results basically due to  compensation of Gaussian decay of non-phononic modes by increasing number of phonoinc modes. This however survives for a very small $\omega$-range beyond $\omega_{ir}$ beyond which the typical Debye behavior of DOS starts dominating.

Besides theoretical formulation of the DOS, the present work also provides an additional insight.
The Hamiltonain $H$ of the superblock is a sparse matrix in the basis space consisting of the product states of the eigenfunctions of the basic blocks. It is therefore expected to go to a many body localized phase  below a system specific temperature. As this phase is believed to violate thermalization in case of an isolated system, it is also referred as a quantum "glass" phase. This hints that an analysis of superblock Hamiltonain in NI basis of basic blocks may help us gain some  insights in glass-transition phenomenon.

In the end, It is worth indicating some connections with other complex systems.  The DOS described in this paper is analogous to that of  the many body DOS for the ensembles of $m$ fermions with $k$-body interactions, also known as embedded ensembles, and very successful in modelling the density of states of interacting fermion systems. 
Another point worth noting here is following: although the Gaussian density  of states here is derived in context of amorphous materials, the approach as well as the result is applicable for  any sparse matrix with similar statistics of the matrix elements.

\acknowledgments

I am grateful to SERB, DST, India for the financial support provided for the  research.

\appendix

\section{Calculation of  $\langle \left( V_{kl} \right)^2 \rangle$  }

\begin{eqnarray}
\langle \left( V_{kl} \right)^2 \rangle  = 
{1\over 4} \sum_{s, t; s\not=t} \; \sum_{s', t'; s'\not=t'} \; \sum_{te, te'}   \; D ^{(st)}_{kl} \;  D ^{(s't')}_{kl} \; U^{(st)}_{\alpha \beta \gamma \delta} \; U^{(s't')}_{\alpha' \beta' \gamma' \delta'} \langle \; \Gamma^{(s)}_{\alpha \beta; kl}  \; \Gamma^{(t)}_{\gamma \delta; kl} \; \Gamma^{(s')}_{\alpha' \beta'; kl}  \; \; \Gamma^{(t')}_{\gamma '\delta'; kl} \rangle
\nonumber \\
\label{vvar0}
\end{eqnarray}
where $\sum_{te, te'} \equiv \sum_{\alpha \beta \gamma \delta} \; \sum_{\alpha' \beta' \gamma' \delta'}$.
Now as the stress-matrix elements of different blocks are not correlated, we can write
\begin{eqnarray}
& &\langle \; \Gamma^{(s)}_{\alpha \beta; kl}  \; \Gamma^{(s')}_{\alpha' \beta'; kl} \; \Gamma^{(t)}_{\gamma \delta; kl}  \; \; \Gamma^{(t')}_{\gamma '\delta'; kl} \rangle \nonumber \\
&= &  2 \langle \; \Gamma^{(s)}_{\alpha \beta; kl}  \; \Gamma^{(s')}_{\alpha' \beta'; kl} \;  \rangle  . \langle  \; \Gamma^{(t)}_{\gamma \delta; kl}  \; \; \Gamma^{(t')}_{\gamma '\delta'; kl} \; \rangle  \\
&=& 2 \left( {N^{-1} \; \omega_u \; \rho_m \; v_a^2  \; \Omega_b}\right)^2 \; \langle Q^{-1}_{0;s} (\omega)\rangle_{e,\omega} \;  \langle Q^{-1}_{0;t} (\omega)\rangle_{e,\omega} \;  
 \; \delta_{ss'} \; \delta_{tt'}  
\label{rel1}
 \end{eqnarray}
where the last relation is obtained by using  eq.(\ref{gcorr3}) and the prefactor $2$ arises from the similar contribution from pairwise combination $\Gamma^{(s)}_{\alpha \beta; kl}  \; \Gamma^{(t')}_{\gamma' \delta'; kl} \;  \rangle  . \langle  \; \Gamma^{(s')}_{\alpha' \beta'; kl}  \; \; \Gamma^{(t)}_{\gamma '\delta; kl} \; \rangle$

Substituting eq.(\ref{rel1}) alongwith eq.(\ref{u}) in eq.(\ref{vvar0}), one gets
\begin{eqnarray}
\langle \left( V_{kl} \right)^2 \rangle
&=& {2 \over (4 \pi \rho_m v_a^2)^2} \; \left( {\omega_u \Omega_b \over N \pi  }\right)^2 \; \sum_{s, t; s\not=t} 
   {\left(D^{(s t)}_{kl}\right)^2 \over |{\bf R_s}-{\bf R_{t}}|^6}   \;  W_{st}
\label{vvar}
\end{eqnarray}
 with 
\begin{eqnarray}
W_{st}  &=&  (\rho_m \; v_a^2)^2  \sum_{te, te'} \;
\langle \; \kappa^{(st)}_{te}   
\;  \kappa^{(st)}_{te'}  \; \rangle \; \; \; Y_{\alpha \beta \alpha' \beta'}  \;
Y_{\gamma \delta \gamma' \delta'} \;
\langle Q^{-1}_{0;s} (\omega)\rangle \;  \langle Q^{-1}_{0;t} (\omega)\rangle 
\label{w}
\end{eqnarray}

As the ensemble as well as energy avearged attenuation constants of two different blocks can safely be assumed to be equal, $W_{st}$ can be written as

\begin{eqnarray}
W_{st} 
&=&(\rho_m v_a^2)^2 \;  K^2 \; \langle Q_0^{-1} ( \omega) \rangle_{\omega} ^2 
\label{w1}
\end{eqnarray}
where
\begin{eqnarray}
K^2= \sum_{te, te'}  \; Y_{\alpha \beta \alpha' \beta'}  \;
Y_{\gamma \delta \gamma' \delta'} \;
\langle \; \kappa^{(st)}_{te}   \;  \kappa^{(st)}_{te'}  \; \rangle
\label{K}
\end{eqnarray} 
with $Y_{\alpha \beta \alpha' \beta'} = q \; \delta_{\alpha \beta} \delta_{\alpha' \beta'} +
\delta_{\alpha \alpha'} \delta_{\beta \beta'}  + \delta_{\alpha \beta'} \delta_{\beta \alpha'} $. Here $\rho$ is the mass-density of a sub-block and $v$ is the speed of sound waves.

Substitution of eq.(\ref{w1}) in eq.(\ref{vvar}) gives
\begin{eqnarray}
\langle \left( V_{kl} \right)^2 \rangle
&=& 2\left( {N^{-1} \; \omega_u  \; K }\right)^2 \; \; \langle Q_0^{-1} ( \omega) \rangle_{\omega} ^2 \;\; C_{kl}
\label{vvar1p}
\end{eqnarray}
with $C_{kl}$ given by eq.(\ref{ckl}) and 
\begin{eqnarray}
K^2=(8/3)[-3+4s+16 q(q+s+q s -1)] =122.1.
\label{kk}
\end{eqnarray}
Here the $2nd$ equality above follows from the typical values of  $q \equiv 1-{ v^2_t \over v^2_l} \simeq 0.6$ and $s \equiv { {\rm Im} \chi_l \over {\rm Im} \chi_t} \simeq 2.6$.

\section{Calculation of $C_{kl}$}

Following from eq.(\ref{dkl}), $D^{(s t)}_{kl}=1$ for all those $(k,l)$ pairs which  differ, if at all,  only in contributions from the  specific blocks $s$ and/or $t$  and is zero otherwise. For example, if $|k \rangle, |l \rangle$ differ in the contributions from a single block, say "p", that is, if $k_p \not= l_p$ with $p \not= s, t$ then $D^{(s t)}_{kl}=0$.  But $D^{(s t)}_{kl}=1$ if $p=s \; {\rm or} \; t$.  
As a consequence, eq.(\ref{ckl}) gives, for a pair $|k \rangle, |l \rangle$ forming a $j$-plet,

\begin{eqnarray}
C_{kl} 
&=& \; \delta_{kl} \; \; {1\over 4\pi^2} \; \sum_{s, t;  s\not=t}  {\Omega_b^2 \over  | \; {\bf R_s}-{\bf R_{t}} \; |^6}   \hspace{0.3in} {\rm for} \; j=0     
\label{cc1}  \\
&=& {1\over 4\pi^2} \;  \sum_t \; {\Omega_b^2 \over  | \; {\bf R_s}-{\bf R_{t}} \; |^6}  
\hspace{0.6in} {\rm for} \; j=1, \; k_s \not=l_s \; {\rm or} \; k_t \not=l_t
\label{cc2}  \\
&=&  {1\over 4\pi^2} \;  {\Omega_b^2 \over  | \; {\bf R_s}-{\bf R_{t}} \; |^6}  
\hspace{0.8in} {\rm for} \; j=2, \; k_s \not=l_s \; {\rm and} \; k_t \not=l_t
\label{cc3}   \\
&=& 0 \hspace{2.0in}  {\rm for} \; \; j>2, 
\label{ckl1}
\end{eqnarray} 
where $"s"$ in eq.(\ref{cc2}) refers to the block whose eigenfunction contribution to $|k\rangle$ and $|l\rangle$ (as they form $1$-plets) is different.  Note the subscripts $k$ and $l$ are not explicitely present on the right side of eq.(\ref{cc2}) and therefore $C_{kl}$ for all $k,l$ pairs which differ in the contribution from the same block "s" will be equal.   Similarly $"s"$ and $"t"$ in eq.(\ref{cc3}) correponds to two blocks whose contributons (eigenfunctions) to
$|k\rangle$ are diiferent from the ones to $|l\rangle$ (as they form $2$-plets). Again the subscripts $k$ and $l$ are absent on right side of eq.(\ref{cc3}) and therefore  $C_{kl}$ for all $k,l$ pairs which differ in the contribution from the same blocks "s,t" will be equal.   

Following from the above, $C_{kl}$ for  $k, l$ pairs   forming $2$-plets are quite small as compared to those forming $0$ and $1$-plets.

As the case of a $k,l$ pair forming a $0$-plet is possible only if $|k\rangle=|l\rangle$, it corresponds to a diagonal matrix element.  This gives the total number of $0$-plets as the size of NI basis i.e $N^g$ with $N$ as the number of single block energy levels and $g$ as the number of basic blocks in the superblock. 
Also all clear from eq.(\ref{cc1}), all $C_{kk}$ are equal i.e $C_{kk}=C_0$ where 
\begin{eqnarray}
C_0 &\equiv & {1\over 4\pi^2} \;   \sum_{s, t;  s\not=t}  \; {\Omega_b^2 \over  | \; {\bf R_s}-{\bf R_{t}} \; |^{6}} \approx  g \; d_0
\label{c00}
\end{eqnarray}
where $d_0$ is a dimensionless constant. As the dominant contribution to $\sum_{s,t}$ comes from the neighboring blocks, $d$ can be estimated as  $d_0 \sim {z\over 144}$ (assuming  basic block of spherical shape  with $\Omega_b = (4 \pi/3) R_b^3$ and $R_b \approx |R_s -R_t|/2$ for two neighboring blocks with $z$ as the number of nearest neighbors of a given block). The above implies
\begin{eqnarray}
\sum_{k,l \; \epsilon \; 0-plet } C_{kl} =  \sum_k C_{kk} = N^g \; C_0
\label{cd0}
\end{eqnarray}

The total number of $k,l$  pairs forming $1$-plets is $g  (N-1) N^g$ and these will arise from all possible $s, t$ pairs chosen out of $g$ basic blocks.  But only a total of $N^g (N-1)$ possible pairs of such eigenvectors $|k \rangle, |l \rangle$ correspond to a same  block $s$; from eq.(\ref{cc3}), $C_{kl}$ for all such $k,l$ pairs is equal. As a consequence, we have

\begin{eqnarray}
\sum_{k,l \; \epsilon \; 1-plet } C_{kl} 
&=& {N^g (N-1) \over 4\pi^2} \;   \sum_{s, t;  s\not=t}  \; {\Omega_b^2 \over  | \; {\bf R_s}-{\bf R_{t}} \; |^6}  = N^g \; (N-1) \; C_0 \nonumber \\
\label{cd2}
\end{eqnarray}

The total number of $|k \rangle, |l \rangle$ basis pairs forming $2$-plets is ${g(g-1)\over 2} (N-1)^2 N^g$, arising from all possible basic block pairs $s,t$ but only a total of $N^g \; (N-1)^2$ pairs of them correspond to a same pair of blocks $s,t$; from eq.(\ref{cc3}), $C_{kl}$ for all such $k,l$ pairs, which arise from same block-pair $s,t$, is equal. As a consequence, we have

\begin{eqnarray}
\sum_{k,l \; \epsilon \; 2-plet } C_{kl} 
&=& {N^g (N-1)^2 \over 4\pi^2} \;   \sum_{s, t;  s\not=t}  \; {\Omega_b^2 \over  | \; {\bf R_s}-{\bf R_{t}} \; |^6}  = N^g \; (N-1)^2 \; C_0  \nonumber \\
\label{cd1}
\end{eqnarray}

From, eqs.(\ref{cd0}, \ref{cd1}, \ref{cd2}), we have

\begin{eqnarray}
\sum_{kl} C_{kl} &=&
\sum_{k} C_{kk} + \sum_{k,l \; \epsilon \; (g-1)-plet } C_{kl} +\sum_{k,l \; \epsilon \; (g-2)-plet } C_{kl} \nonumber \\
&=&  \left[1+ (N-1)+ (N-1)^2 \right] \; N^g \; C_0 \nonumber \\
&\approx &  N^{g+2} \; \; C_0
\label{cd3}
\end{eqnarray}

\section{  Calculation of  $\langle \left( V_{kl} \right)^n \rangle $}

\begin{eqnarray}
\langle \left( V_{kl} \right)^n \rangle  = 
\sum_S \; \sum_{te_1 \ldots te_n}    \left( \prod_{p=1}^n D^{(s_p t_p)}_{kl}  \; U^{(s_p t_p)}_{\alpha_p \beta_p \gamma_p \delta_p}  \right) \; F_{st}
\label{nvar0}
\end{eqnarray}
where $\sum_S \equiv \sum_{s_1, t_1\atop s_1\not=t_1} \ldots \sum_{s_n, t_n \atop s_n\not=t_n}$ and
$\sum_{te_1\ldots te_n} \equiv \sum_{\alpha_1 \beta_1 \gamma_1 \delta_1} \; \ldots \sum_{\alpha_n \beta_n \gamma_n \delta_n}$ and 

\begin{eqnarray}
F_{st}  =\langle \; \Gamma^{(s_1)}_{\alpha_1 \beta_1; kl}  \; \Gamma^{(t_1)}_{\gamma_1 \delta_1; kl} \; \Gamma^{(s_2)}_{\alpha_2 \beta_2; kl}  \; \Gamma^{(t_2)}_{\gamma_2 \delta_2; kl} \ldots \Gamma^{(s_n)}_{\alpha_n \beta_n; kl}  \; \Gamma^{(t_n)}_{\gamma_n \delta_n; kl}  \rangle 
\label{nvar1}
\end{eqnarray}

Here again, the pairwise correlations between the stress-matrix elements of different blocks are negligible and only those between same blocks are relevant.  For example, for $n=2m$, $F_{st}$ becomes
\begin{eqnarray}
F_{st} &= &  \prod_{i,j=1}^{2m} \langle \; \Gamma^{(s_i)}_{\alpha_i \beta_i; kl}  \; \Gamma^{(s_j)}_{\alpha_j \beta_j; kl} \;  \rangle  . \langle  \; \Gamma^{(t_i)}_{\gamma_i \delta_i; kl}  \; \; \Gamma^{(t_j)}_{\gamma_j\delta_j; kl} \; \rangle  \\
 &=& \left( {N^{-1} \; \omega_u \; \rho_m \; v_a^2 \; \Omega_b }\right)^{2m} \; \prod_{i,j=1}^{2m}  \left( \langle Q^{-1}_{0;s_i} (\omega)\rangle_{\omega} \;  \langle Q^{-1}_{0;t_i} (\omega)\rangle_{\omega}  \; \delta_{s_i s_j} \; \delta_{t_i t_j}  \right) 
\label{nvar2}
 \end{eqnarray}
where the $2nd$ equality is obtained by using eq.(\ref{gcorr3}) respectively. 

The presence of terms $\delta_{s_i s_j} \delta_{t_i t_j}$ in eq.(\ref{nvar2}) ensures contribution to  $\langle \left( V_{kl} \right)^{2m} \rangle$  only from those terms in the sum $\sum^S$ which correspond to  pairwise combinations among $s_1 s_2 \ldots s_{2m}$ (and among $t_1 t_2 \ldots t_{2m}$). In absence of any other condition (except $s_i \not=t_i$) on pairwise combinations, the number of such combinations can be $^{2m} C_2 = (2m-1)!!$. Thus one can replace  $\sum^S$ by the sum over even indices only:
$\sum^S = (2m-1)!! \;\sum_{s_2, t_2, s_4, t_4, \ldots, s_{2m}, t_{2m} \atop s_2\not=t_2, s_4 \not=t_4, \ldots, s_{2m} \not=t_{2m}}$.
 As each such combination has same contribution, one can write
\begin{eqnarray}
\langle \left( V_{kl} \right)^{2m} \rangle  
&=& (2m-1)!! \; \left( {\omega_u \; \Omega_b  \over \pi  N}\right)^{2m} \sum_{s_2, t_2, s_4, t_4, \ldots, s_{2m}, t_{2m} \atop s_2\not=t_2, s_4 \not=t_4, \ldots, s_{2m} \not=t_{2m}}
   \prod_{p=2}^{2m} \; {D^{(s_p t_p)}_{kl} \; W_{s_p t_p} \over ({\bf R_{s_p}}-{\bf R_{t_p}})^6 } 
\label{nvar3}
\end{eqnarray}
 with $W_{s_p t_p}$ is same as eq.(\ref{w}).

The summation and product over terms in the above equation can  be rearranged to write
\begin{eqnarray}
\langle \left( V_{kl} \right)^{2m} \rangle  
&=& (2m-1)!! \; \left( {\omega_u \; \Omega_b \over \pi N }\right)^{2m} \; \prod_{p=2}^{2m} \left( \sum_{s_p, t_p \atop s_p\not=t_p}
 \; {D^{(s_p t_p)}_{kl} \; W_{s_p t_p} \over ({\bf R_{s_p}}-{\bf R_{t_p}})^6 } \right)
\label{nvar4}
\end{eqnarray}

Substitution of eq.(\ref{w1}) and eq.(\ref{K}) in eq.(\ref{nvar3}) gives
\begin{eqnarray}
\langle \left( V_{kl} \right)^{2m} \rangle  
&=& (2m-1)!! \; \left( {\omega_u \; \Omega_b  \over \pi  N}\right)^{2m} \; \; \langle Q_0^{-1} ( \omega) \rangle_{\omega} ^{2m} \; 
\prod_{p=2}^{2m} \left( \sum_{s_p, t_p \atop s_p\not=t_p}
 \; {D^{(s_p t_p)}_{kl} \over ({\bf R_{s_p}}-{\bf R_{t_p}})^6 } \right) \\
 &=& (2m-1)!! \; \left( {\omega_u \; \Omega_b  \over \pi  N}\right)^{2m} \; \; \langle Q_0^{-1} ( \omega) \rangle_{\omega} ^{2m} \; \left(C_{kl} \right)^{2m} \\
&=& (2m-1)!! \;   \langle \left( V_{kl} \right)^2 \rangle^m 
\label{nvar5}
\end{eqnarray}
which is same as eq.(\ref{l2n}).

\newpage

\begin{table}[ht!]
\caption{\label{tab:table II} {\bf Comparison of Theoretically predicted range and location of excess VDOS with experimental data} ({\it all energy scales in {\rm THz} units}): Based on inelastic neutron and Raman scattering, the  studies \cite{ya26, ya21, ya27, ya28} describe the experimental behavior for excess VDOS for five materials, namely, Selenium, $B_2O_3$, orthoterphenayl, Polybutadiene (PB), and Glycerol in the energy range upto $6-8 \; {\rm mev}$ ($1 \; {\rm THz}= 4.136 \;  {\rm meV} = 6.64 \times 10^{-22}\; {\rm Joules}$). Using the {\it Plotdigitzer} software, we digitally scan the figures in \cite{ya26, ya21, ya27, ya28} and display the experimental data  for total DOS as well as excess DOS in figures 1-7 along with  a numerical fitting in each case (using standard GNU software). This table  gives the fitted and theoretical values of all the relevant energy scales  in case of excess DOS (parts (b) of figures 1-7) i.e $e_d$ (Debye, eq.(\ref{debye})), $e_{ir}$ (Ioffe-Regel, eq.(\ref{eu})), $e_{bp}$ (boson peak, eq.(\ref{mu})), $\sigma$ (variance, eq.(\ref{sig3})), $e_{gl}$ and $e_{gh}$ (lower eq.(\ref{egl}) and upper limits eq.(\ref{xgh}) of Gaussian VDOS). We note that  $R_b$ in this case can not be determined from eq.(\ref{ve0}) (due to unavailability of the experiemental data for the phonon mediated coupling strengths $\gamma_a$). As the only alternative available left, we use the theoreticaly predicted value for $R_b$ i.e $R_b=R_0=4 R_m$ with $R_m$ as the molecular radius: $R_m ={M \over \rho_m N_a}$ with $M, \rho, N_a$ as the mass of the basic structural unit, mass density and Avogrado number\cite{bb3}. The values of $v_l, v_t$ and $\rho_m$ for Se, PB, B$_2$O$_3$, SiO$_2$ are taken from \cite{mb,pohl} and $M$ from \cite{bb2}. But in case of OTP (Orthoterphenyl), $v_l, v_t$ are taken from \cite{hw} and  $M, \rho_m$ from {\it PubChem}. For Glycerine, $v_l, v_t$ are taken from \cite{lya}. 
Here the deviations of theoretical predictions from experiments in some cases can be attributed to lack of exact information about the  the basic structural unit participating in the dispersion intercations within a basic block and thereby its mass $M$ \cite{bb1,bb2}}.
 \begin{center}
\begin{ruledtabular}
\begin{tabular}{|l|c|ccccc|cc|cc|cc|cc|cc|cr|}
Ind. & Glass  & {$M\atop {gm \over mol}$} & ${\rho \atop \; {gm\over cm^3}}$ & ${R_m \atop \AA}$ & ${v_l \atop {km\over sec}} $ & ${v_l \atop {km\over sec}} $ & ${e_d \atop th.}$ & ${ e_d \atop exp}$ & ${e_{ir} \atop th.}$ & ${ e_{ir} \atop fit}$ & ${\sigma \atop th.} $ & ${\sigma \atop fit}$ & ${e_{bp} \atop th.}$  &  ${e_{bp}\atop fit} $ & ${e_{gl} \atop th.}$ & ${e_{gl} \atop fit}$ & ${e_{gh} \atop th.}$ & ${e_{gh} \atop fit}$   \\
\hline
1  & Se &  78.96 & 4.30 &  1.94 &  2.00 &  1.05 &  3.66  & 3.68 & 1.83 & 1.6  &  0.31 & 0.27 & 0.92 & 0.66 & 0.48 &  0.36 & 1.75 &  1.1  \\
2 & PB &  55.15 & 0.93 &  2.86 &  3.02 &   1.46 &  3.46 & 3.40 & 1.73 & 1.69 & 0.29 &  0.31 & 0.87 & 0.87 & 0.46 &  0.43 & 1.47 &  1.45 \\
3  & B$_2$O$_3$ &  69.62 & 1.80 &  2.48 &  3.47 &  1.91 &  5.18  & 8.07 & 2.59 & 3.87 & 0.43 & 0.95 & 1.30  & 1.81 & 0.69 &  0.68 & 2.46 & 2.41 \\
4  & Glyc.&  92.09 & 1.42 &  2.95 &  3.52 &  1.85 &  4.23  & 6.77 & 2.12 & 2.66 & 0.35 & 0.38 & 1.06  & 1.38 & 0.57 &   -  & 2.01 &  2.01 \\
5  & OTP &  230.3 & 1.16 &  1.99 &  2.94 &  1.37 &  4.69  & 4.35 & 2.35 & 1.2  &  0.39 & 0.25 & 1.17 & 0.58 & 0.62 &  0.34 & 1.03 &  0.9 \\
6  &SiO$_2$ &  120.09 & 2.20 &  2.51   &   5.80 &  3.80 &  10.06 & 15.11 & 5.03 & 4.0 &  0.84 & 0.78 & 2.52 & 1.85 & 1.33 &  0.91 & 4.9 &  2.2  \\
\end{tabular}
\end{ruledtabular}
\end{center}
\end{table}

\newpage

\newpage

\newpage

\begin{figure}[ht!]
\centering

\vspace{-2in}

\includegraphics[width=12 cm,height=16 cm]{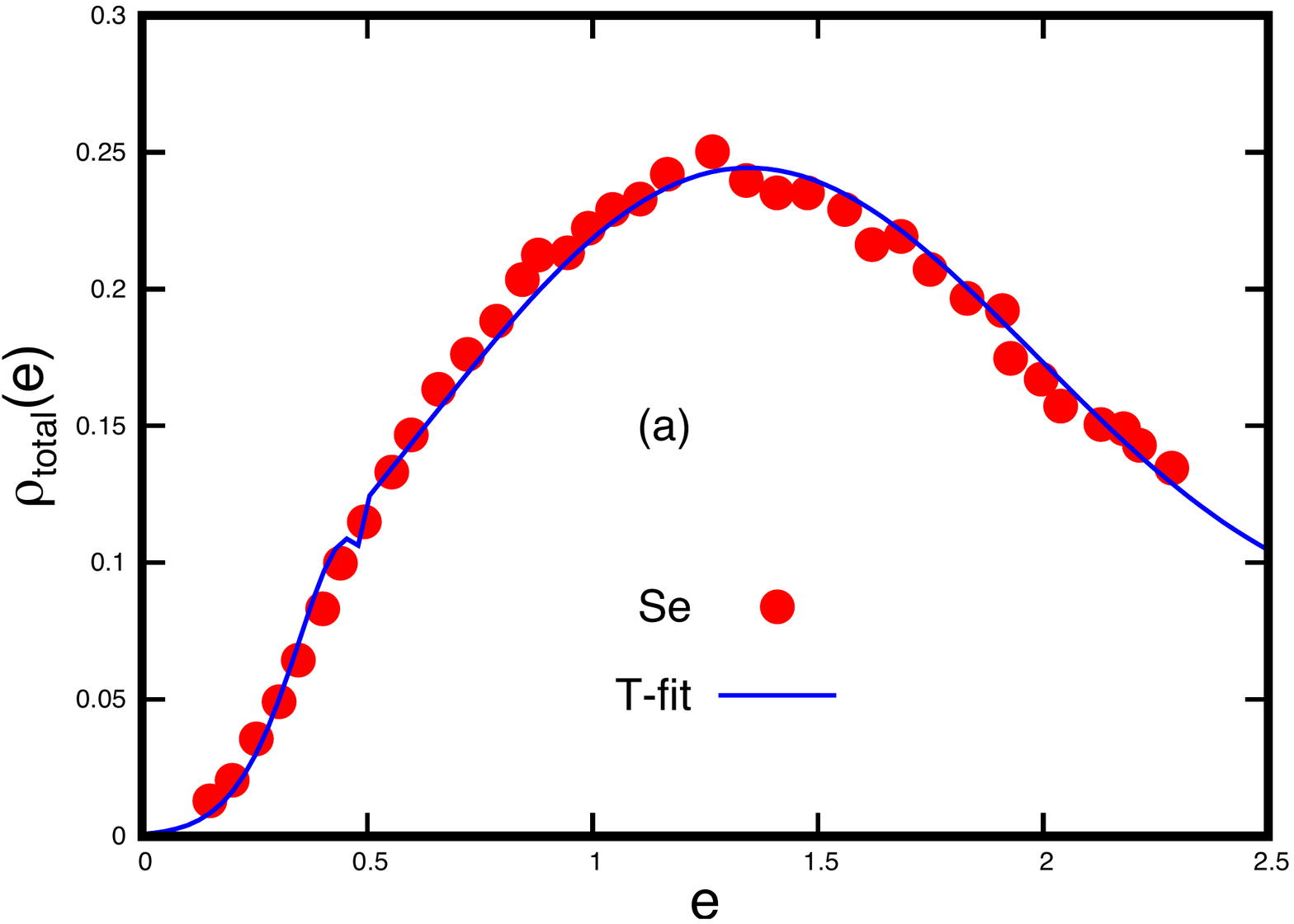}

\vspace{-3.2in}

 \includegraphics[width=12 cm,height=16 cm]{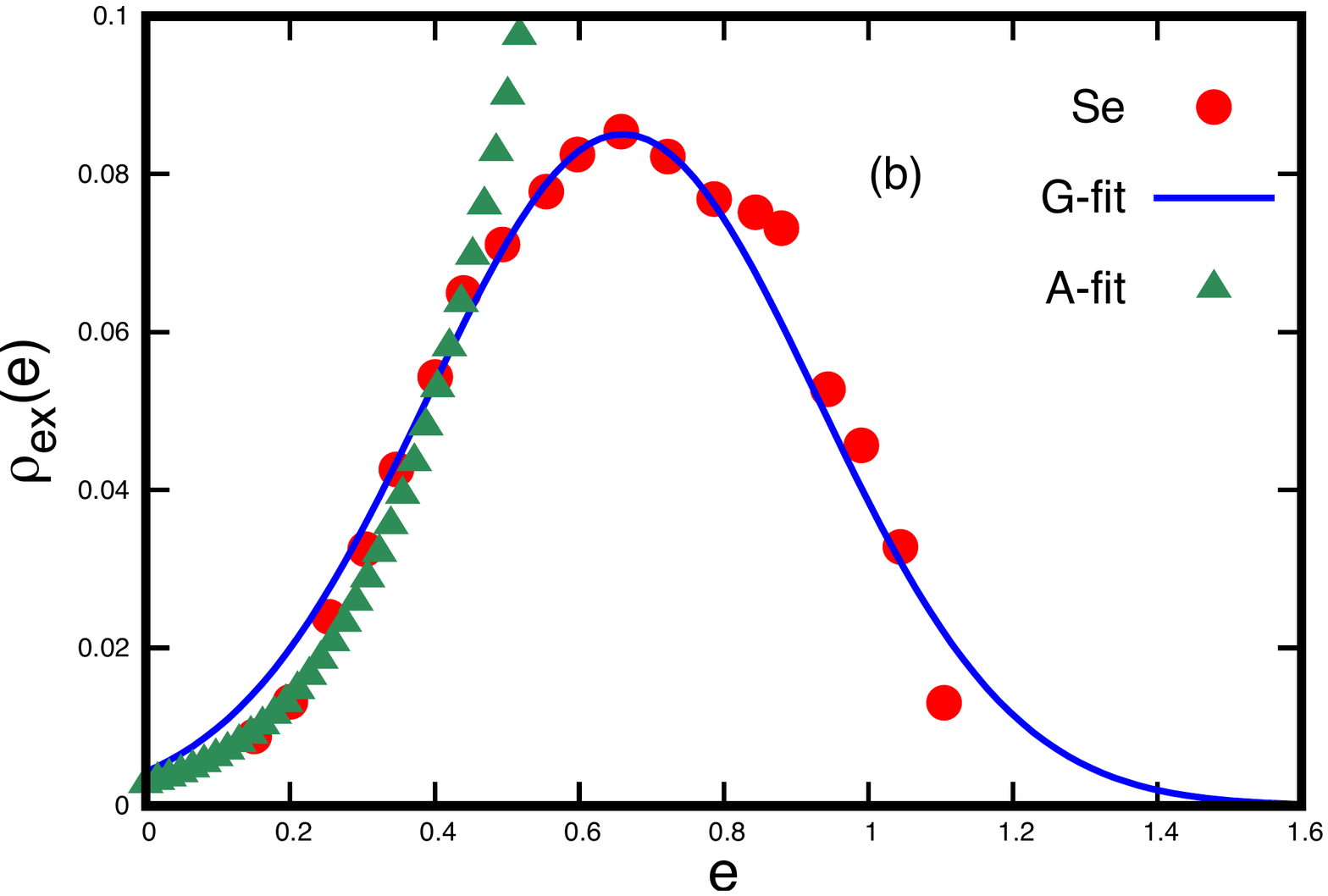}

\vspace{-1.7in}

\caption{{\bf Vibrational DOS for Se} (with $e$ in THz units): The  figure illustrates a comparison of theoretical prediction (solid line) with experimental data (filled circles) (latter obtained by a digital scan of the figure 3 in \cite{ya25} for the total DOS using the {\it Plotdigitzer} software): (a) total vibrational DOS $\rho_{total} \equiv \langle \rho_{total}(e) \rangle$ with $e=\hbar \omega$, (b) excess vibrational DOS $\rho_{ex} \equiv \langle \rho_{total}(e) \rangle -\rho_{Debye}(e)$ where $\rho_{Debye}(e) = {\mathcal D}_b \; e^2$.
The part (a) shows a theoretical fitting of eq.(\ref{gau4})  (''T-fit'' = $0.2 \; {\rm Ai} \left[-9.5 (e-0.35) \right] \; (1-\Theta(e-0.5)) +{\Theta(e-0.5)\over \sqrt{6 \pi}}  {\rm exp}\left[-{(e-1.3)^2\over 2 (0.5)} \right] +0.008 e^2$).  Here $\Theta(x-a)$ refers to Heavyside step function: $\Theta(x-a)=0$ for $x <a$ and $\Theta(x-a) =1$ for $x>a$. The part (b) displays two fits i.e a Gaussian  (''G-fit'' = ${1\over \sqrt{44 \pi}}  {\rm exp}\left[-{ (e-0.66)^2\over 2 (0.073)}\right]$ and an Airy function (''A-fit''=$0.5 \; {\rm Ai} \left[-4.7 (e-0.66) \right]$ ) separately (to verify that the DOS in the lower edge indeed behaves as an Airy function).   We note that the fitting parameters for the case (a) are different from that of (b). This is because while in part (a),  ${\mathcal D}_b=0.008 \; {THz}^{-3}$,  is obtained by fitting the experimental data,  the excess DOS in part (b) is obtained by substracting the Debye DOS data given in figure 3 of \cite{ya25} (the experimental data in the latter gives ${\mathcal D}_b=0.18 \; {THz}^{-3}$). }
 \end{figure}

\begin{figure}[ht!]
\centering

\vspace{-1in}

\includegraphics[width=12cm,height=16 cm]{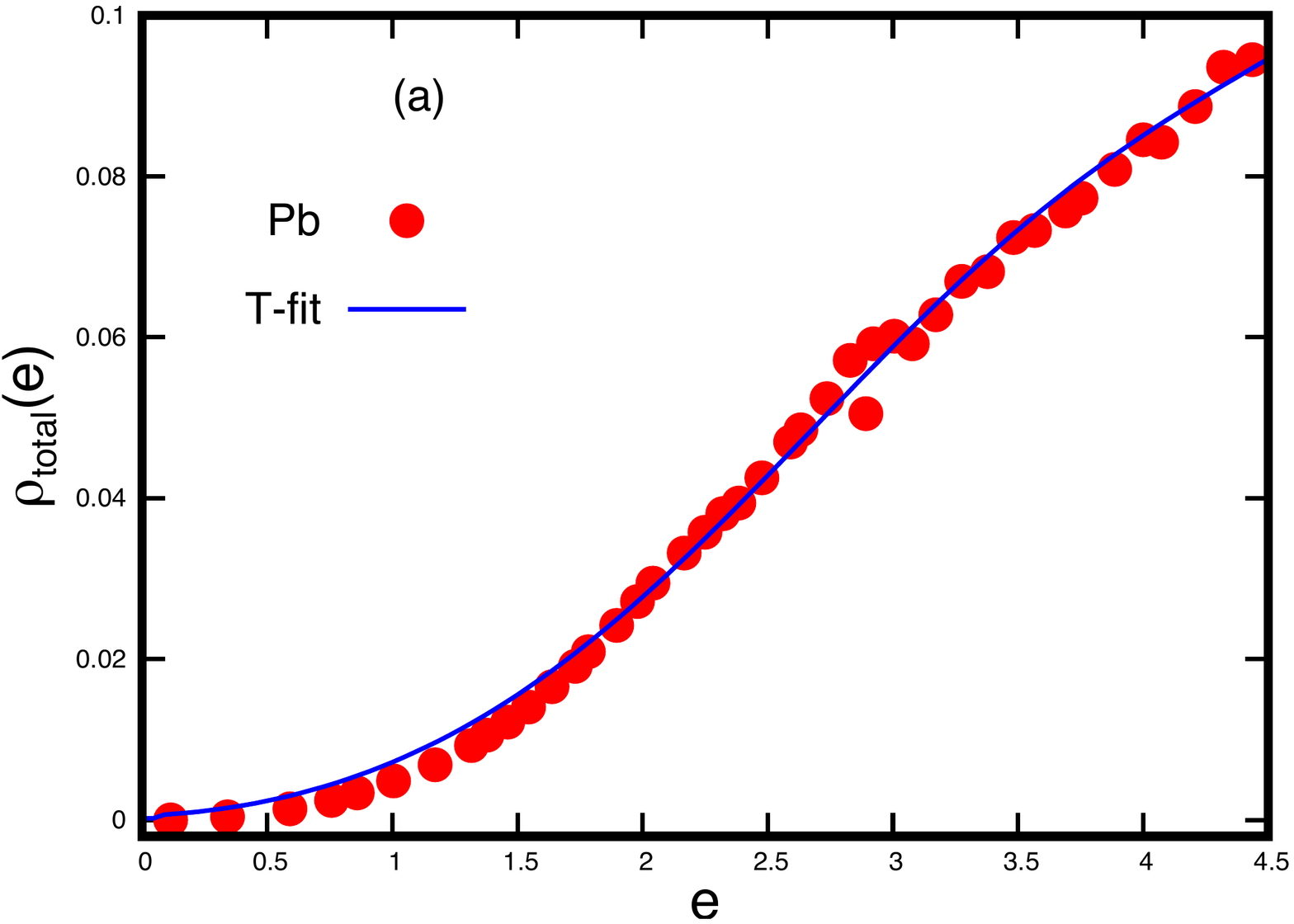}

\vspace{-3in}

\includegraphics[width=12cm,height=16cm]{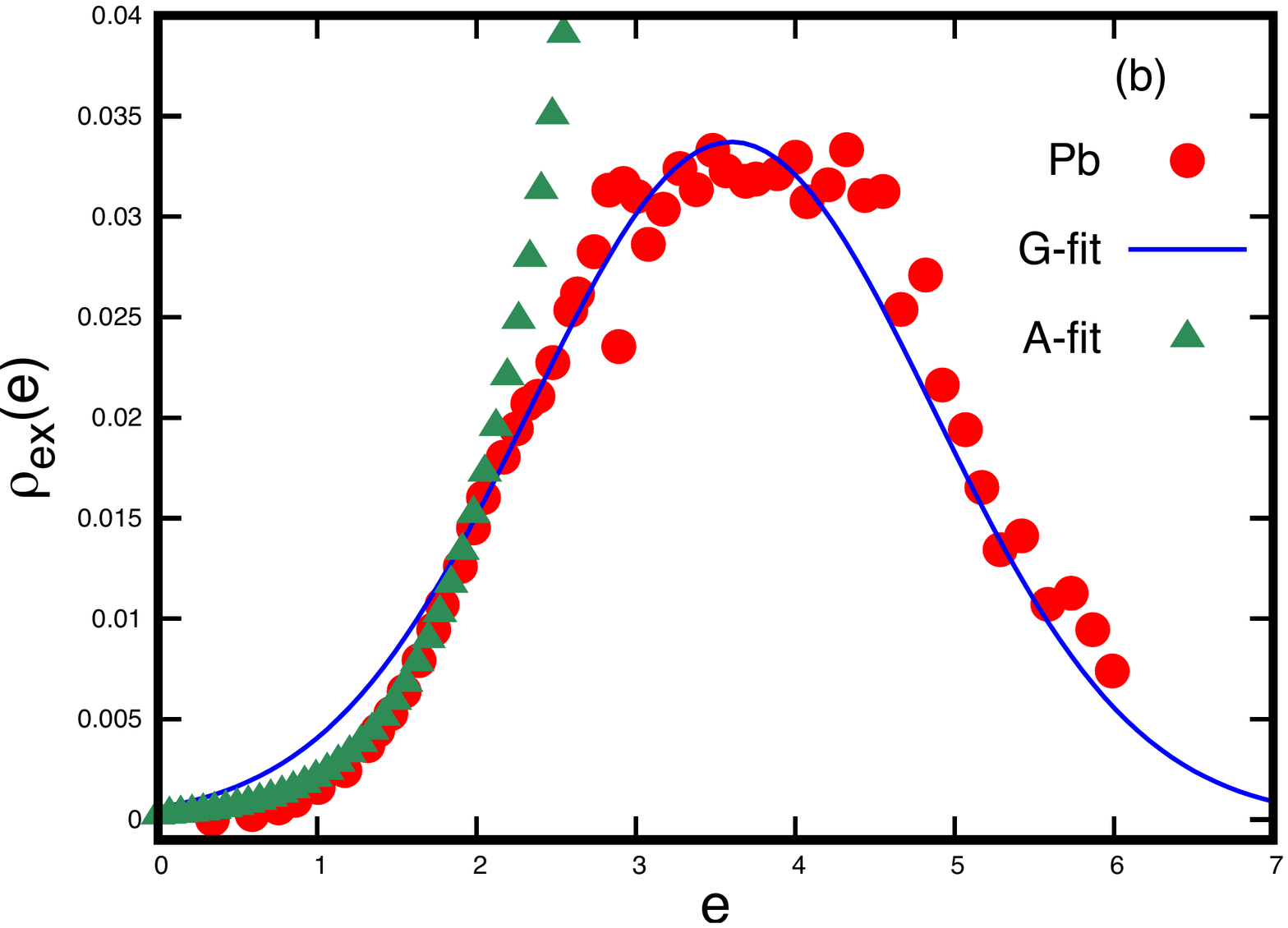}

\vspace{-1.5in}

\caption{{\bf Vibrational DOS for Pb} (with $e$ in mev units):
The experimental data here is obtained by a digital scan of the figure 12 in \cite{ya27} for the reduced DOS ${\langle \rho(e) \rangle\over e^2}$ (with $e$ in $mev$ units) and then using it to obtain  (a) $\langle \rho(e) \rangle$ and (b) $\rho_{excess}=\langle \rho(e) \rangle- {\mathcal D}_b \; e^2$.  Here  ''T-fit''=$0.4 {\rm Ai}(-1.2 (x-3.6)) \; (1-\Theta(x-0.5)) +{\Theta(x-0.5)\over \sqrt{340 \pi}}  {\rm exp} \left[-{ (x-3.6)^2\over 2 (1.6)} \right] +0.0035 \; x^2$ and  ''G-fit''=${1\over \sqrt{280 \pi}}  {\rm exp}\left[-{(x-3.6)^2\over 2 (1.6)}  \right]$ and ''A-fit''=${\rm Ai}(x)=0.4 \; {\rm Ai} \left[-1.2 (e-3.6)\right]$. All other details are same as in figure 2.   We note that here the  fitting parameters for case (a) are in excellent agreement with those of (b). Also note that  fitting of the experimental data in part (a) gives ${\mathcal D}_b=0.0035 \; {mev}^{-3}$ which is very close ${\mathcal D}_b=0.0032 \; {mev}^{-3}$ to that   given by figure 12 of \cite{ya27}.}
\end{figure}

\begin{figure}[ht!]
\centering

\vspace{-1in}

\includegraphics[width=12 cm,height=16 cm]{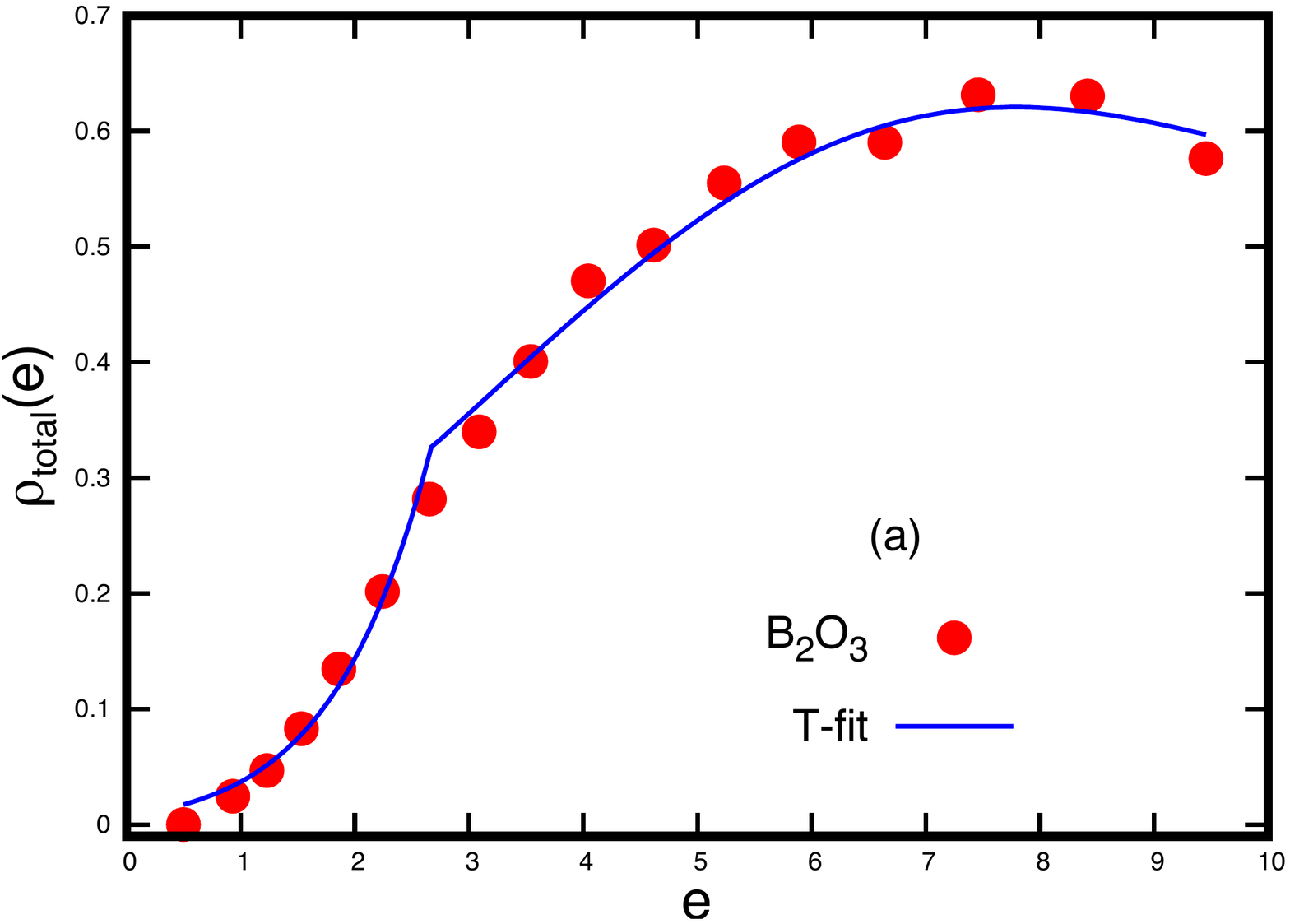}

\vspace{-3in}

\includegraphics[width=12 cm,height=16 cm]{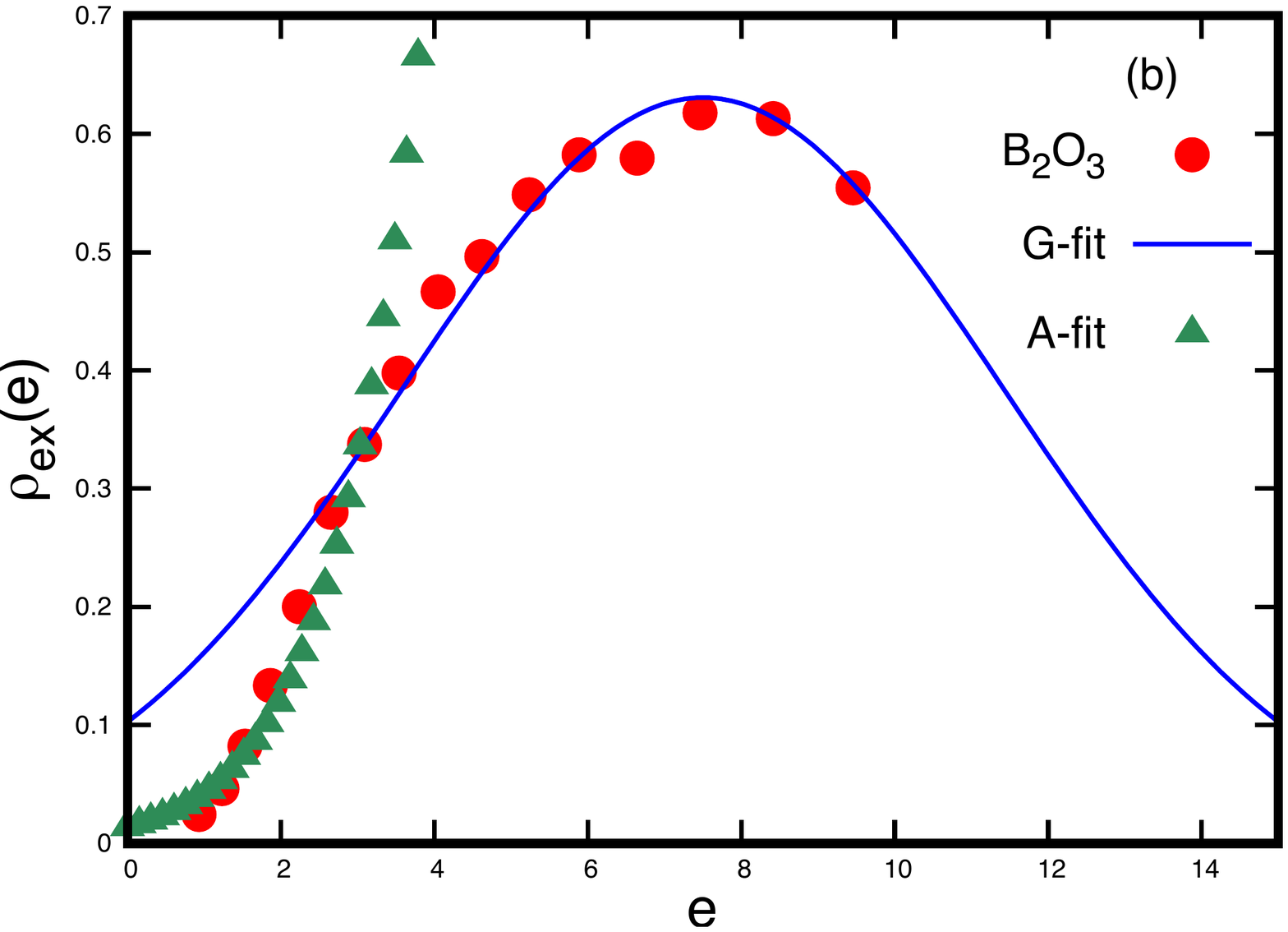}

\vspace{-1.5in}

\caption{{\bf Vibrational DOS for B$_2$O$_3$} (with $e$ in mev units): 
The experimental data here is obtained by a digital scan of the figure 12 in \cite{ya26} for reduced DOS with fits details as follows:  ''T-fit''=$26.7 {\rm Ai}(-0.7 (x-6.5)) \; (1-\Theta(x-2.7)) +{\Theta(x-2.7)\over \sqrt{1.3 \pi}}  {\rm exp} \left[-{ (x-6.5)^2\over 2 (15.5)} \right] +0.0035 \; x^2$, ''G-fit''=${1\over \sqrt{280 \pi}}  {\rm exp}\left[-{(x-7.5)^2\over 2 (15.5)}  \right]$ and ''A-fit''=${\rm Ai}(x)=21.5 \; {\rm Ai} \left[-0.56 (e-7.5)\right]$. All other details are same as in figure 2.   Here again both the Gaussian as well as Airy function fitting for the case (a) is slightly different from that of (b). Further ''T-fit'' in part (a) gives ${\mathcal D}_b=2.9 \times 10^{-4} \; {mev}^{-3}$ which is close ${\mathcal D}_b=2.43 \times 10^{-4}  \; {mev}^{-3}$ to given by figure 4 of \cite{ya28}}.
\end{figure}

\begin{figure}[ht!]
\centering

\vspace{-1in}

\includegraphics[width=12 cm,height=16 cm]{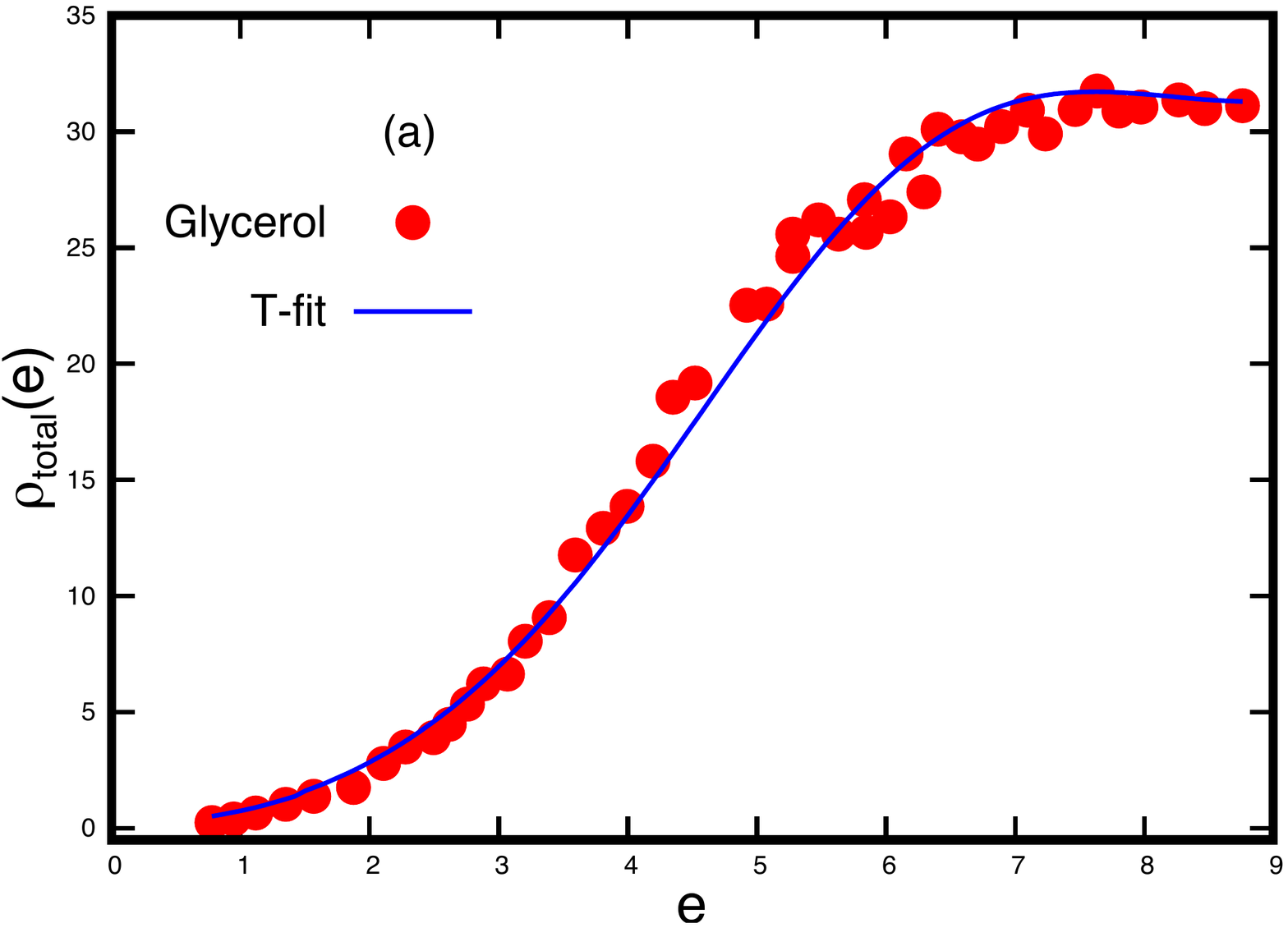}

\vspace{-3in}

\includegraphics[width=12 cm,height=16 cm]{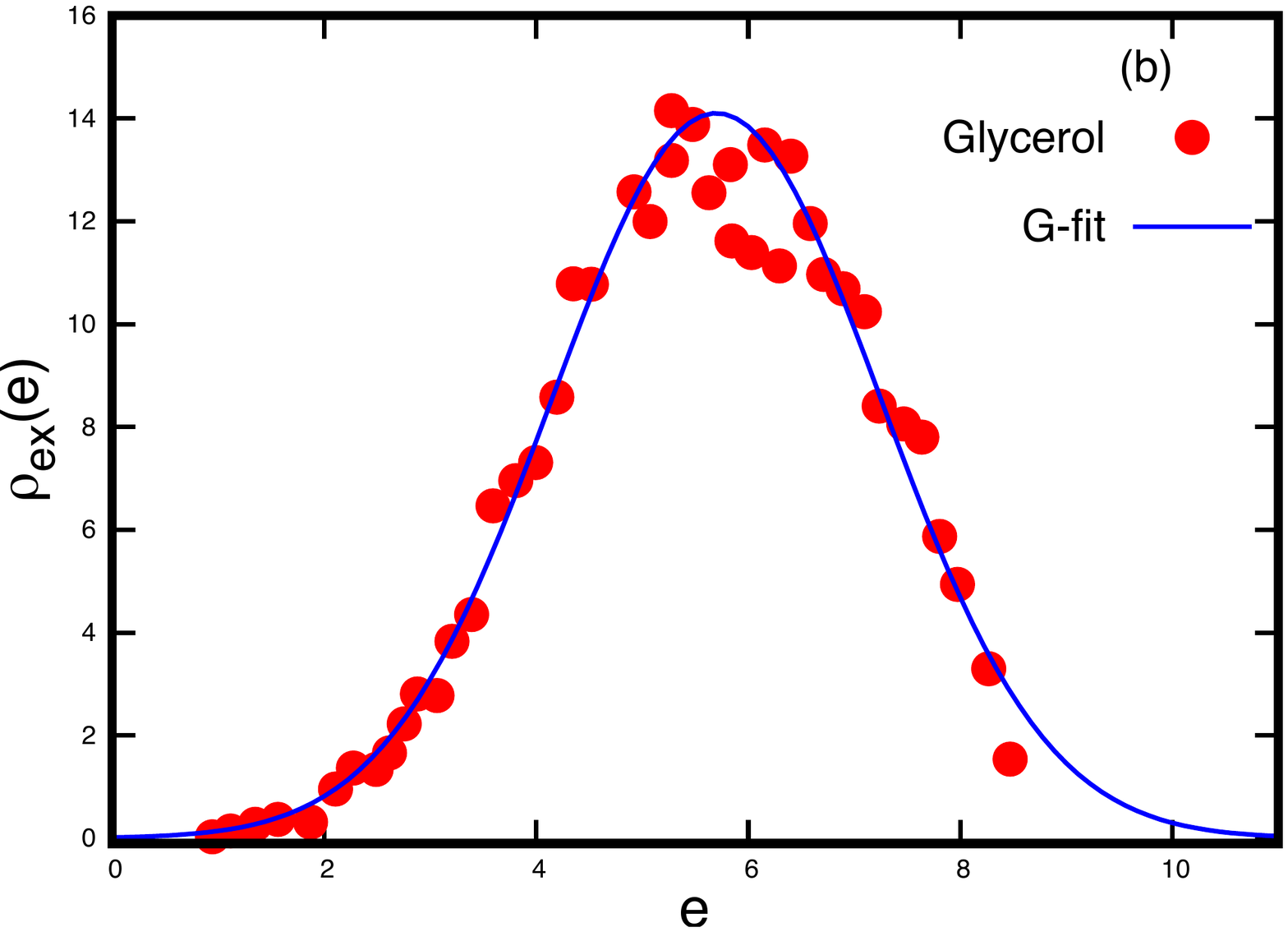}

\vspace{-1.5in}

\caption{{\bf Vibrational DOS for Glycerol} (with $e$ in mev units):  
The experimental data here is obtained by a digital scan of the figure 4 in \cite{ya28} for reduced DOS with fits details as follows:  ''T-fit''=$6.4 {\rm Ai}(-1.0 (x-2.5)) \; (1-\Theta(x-1.5)) +{\Theta(x-1.5)\over \sqrt{340 \pi}}  {\rm exp} \left[-{ (x-6.4)^2\over 2 (4.1)} \right] +0.29 \; x^2$ and  ''G-fit''= ${1\over \sqrt{0.0016 \pi}}  {\rm exp}\left[-{(x-5.7)^2\over 2 (2.4)}  \right]$, referred as ''G-fit''.  All other details are same as in figure 2.  We note here the validity of a Gaussian behaviour in the full range of part (b).   Here again the Gaussian fitting for the case (a) is slightly different from that of (b). Further ''T-fit'' in part (a) gives ${\mathcal D}_b=2.9 \times 10^{-4} \; {mev}^{-3}$ which is close ${\mathcal D}_b=2.43 \times 10^{-4}  \; {mev}^{-3}$ to given by figure 4 of \cite{ya28}}.
\end{figure}

\begin{figure}[ht!]
\centering

\vspace{-1in}

\includegraphics[width=12 cm,height=16 cm]{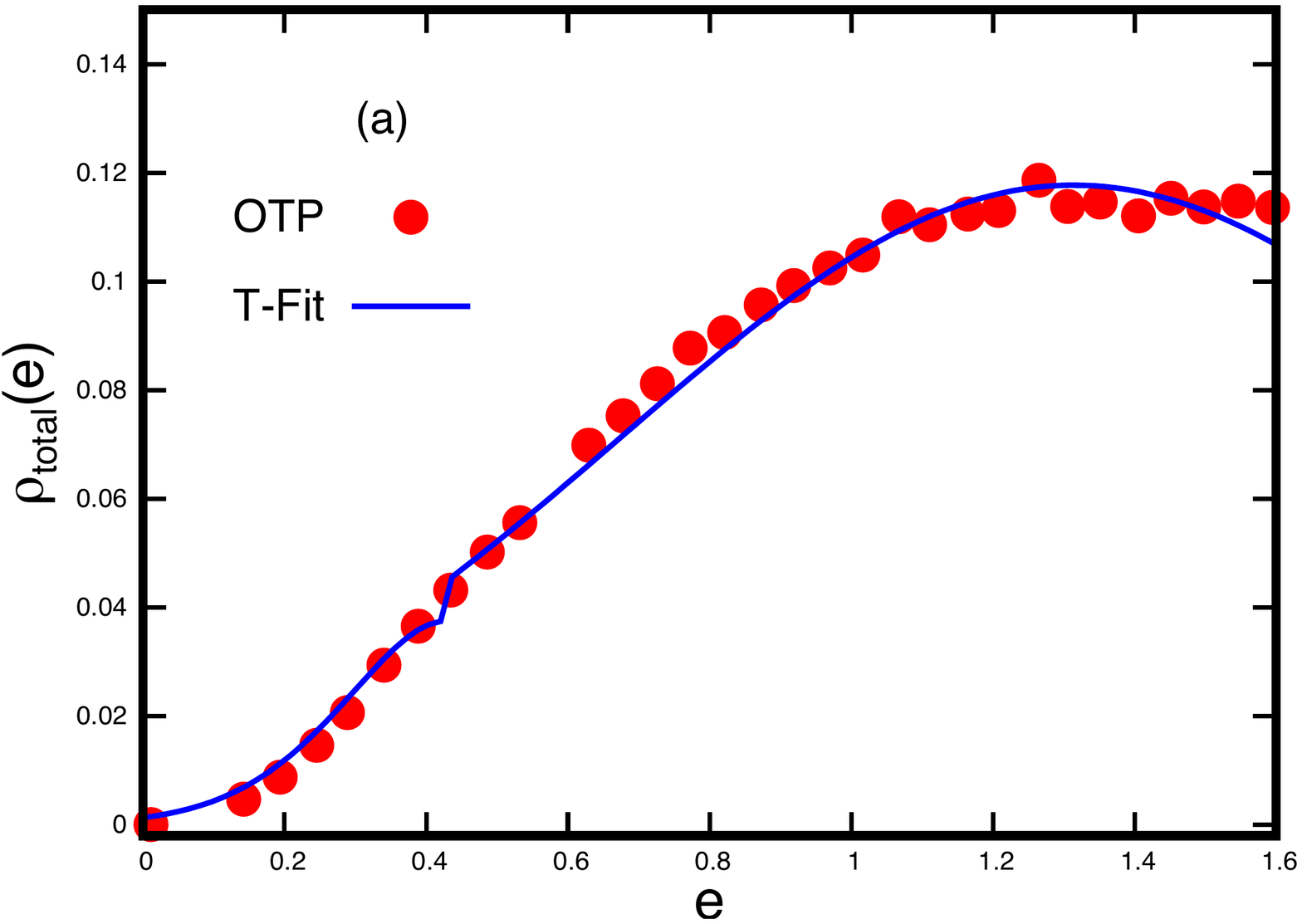}

\vspace{-3in}

\includegraphics[width=12 cm,height=16 cm]{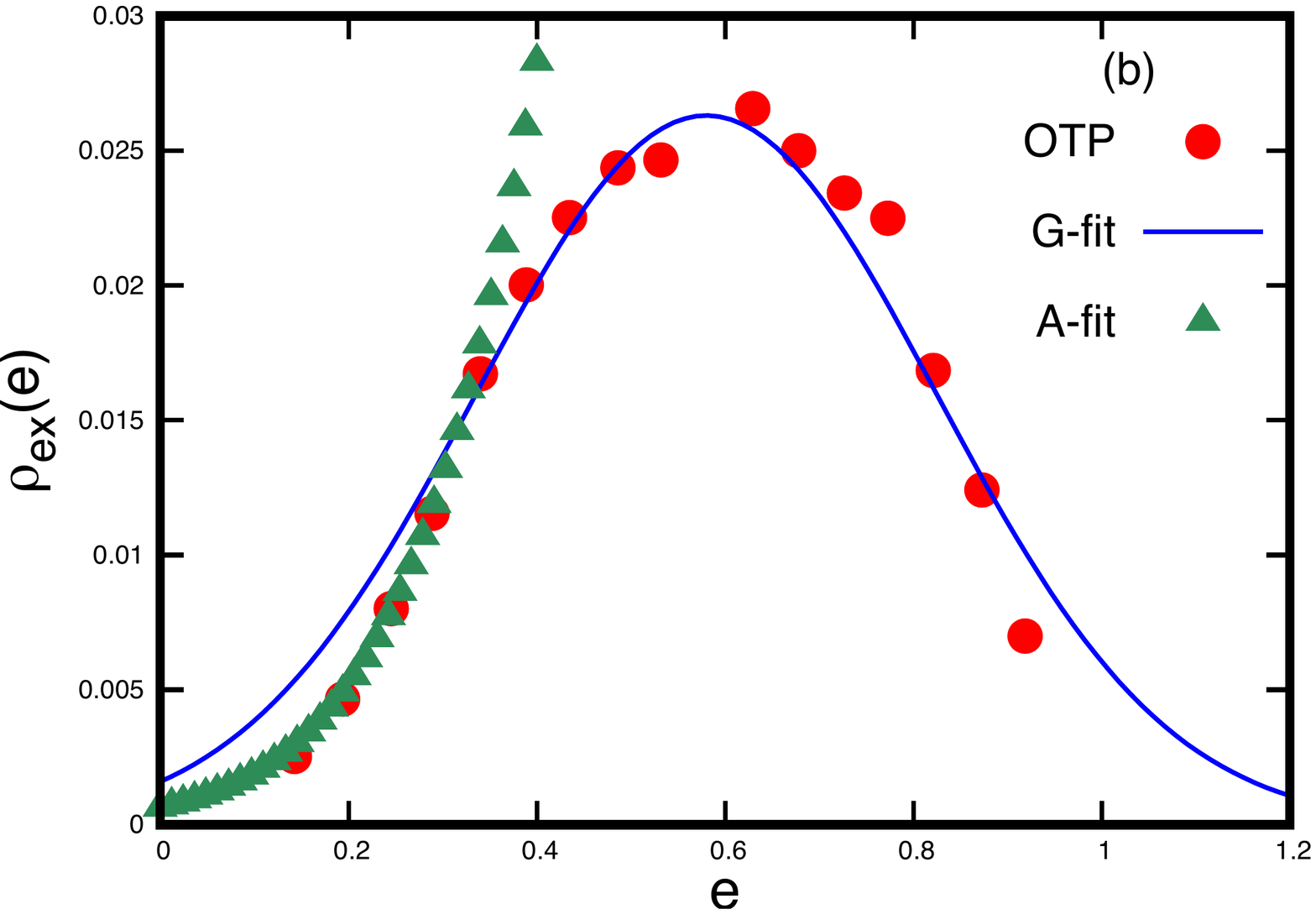}

\vspace{-1.5in}

\caption{{\bf Vibrational DOS for OTP} (with $e$ in THz units):  
The experimental data here is obtained by a digital scan of the figure 3 in \cite{ya21} for the reduced DOS (with $e$ in $THz$ units).  All other details are same as in figure 2 except now  the fits are different: ''T-Fit''=$0.23 {\rm Ai}(-6.0 (x-0.58)) \; (1-\Theta(x-0.43)) +{\Theta(x-0.43)\over \sqrt{24 \pi}}  {\rm exp} \left[-{ (x-1.3)^2\over 2 (0.4)} \right] +0.0015 \; x^2$, ''G-fit''=${1\over \sqrt{460 \pi}}  {\rm exp}\left[-{(x-0.58)^2\over 2 (0.06)}  \right]$ and ''A-fit''=${\rm Ai}(x)=0.23 \; {\rm Ai} \left[-6.0 (e-0.58)\right]$.  Here again both the Gaussian as well as Airy function fitting for the case (a) is slightly different from that of (b). Further Part (a) gives ${\mathcal D}_b \approx 0.0015$ but  the figure 3 of \cite{ya21} gives (${\mathcal D}_b=0.109 \; {THz}^{-3}$)}
\end{figure}

\begin{figure}[ht!]
\centering

\includegraphics[width=14.5cm,height=9 cm]{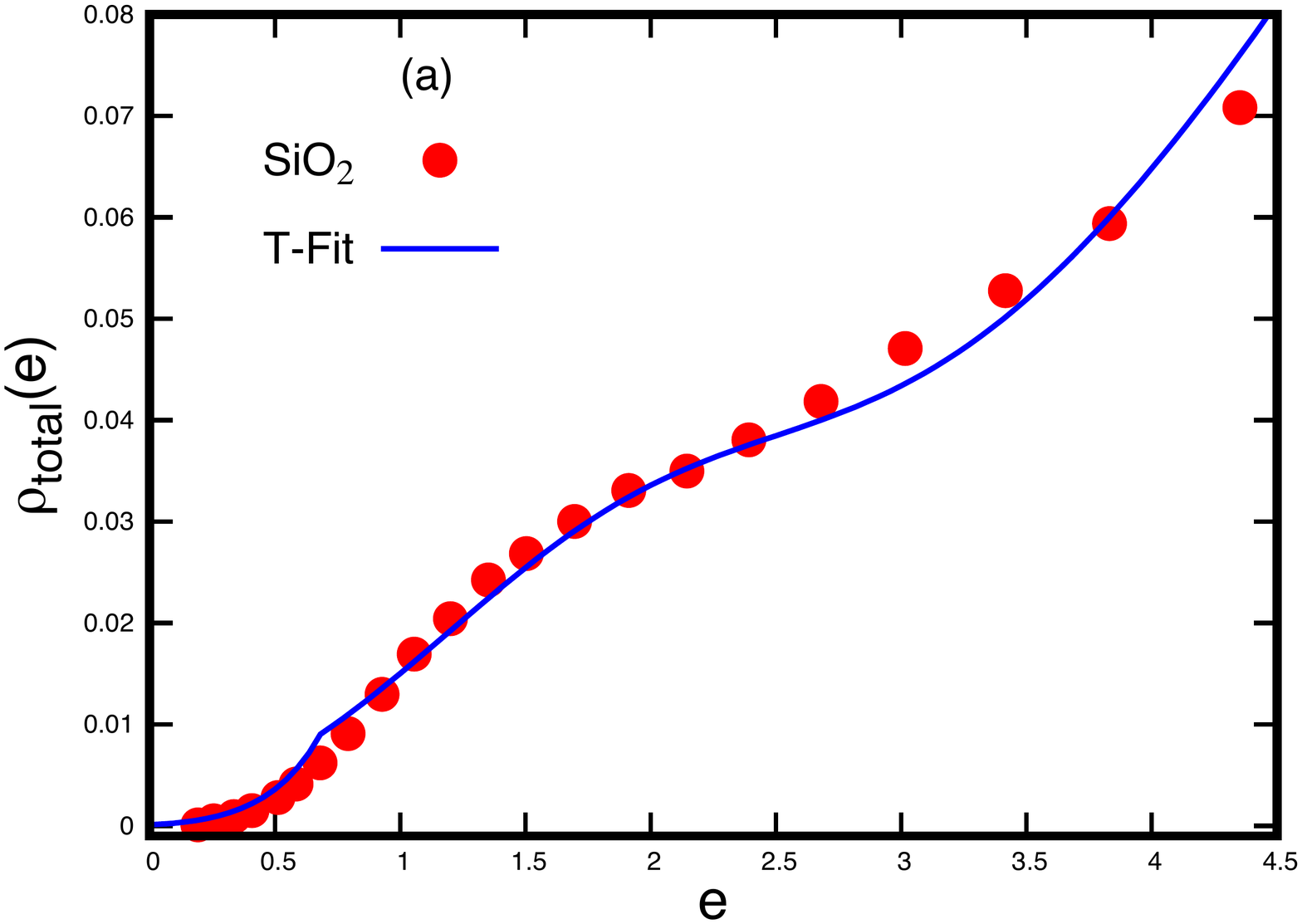}

\vspace{-1.5in}

\includegraphics[width=12cm,height=16 cm]{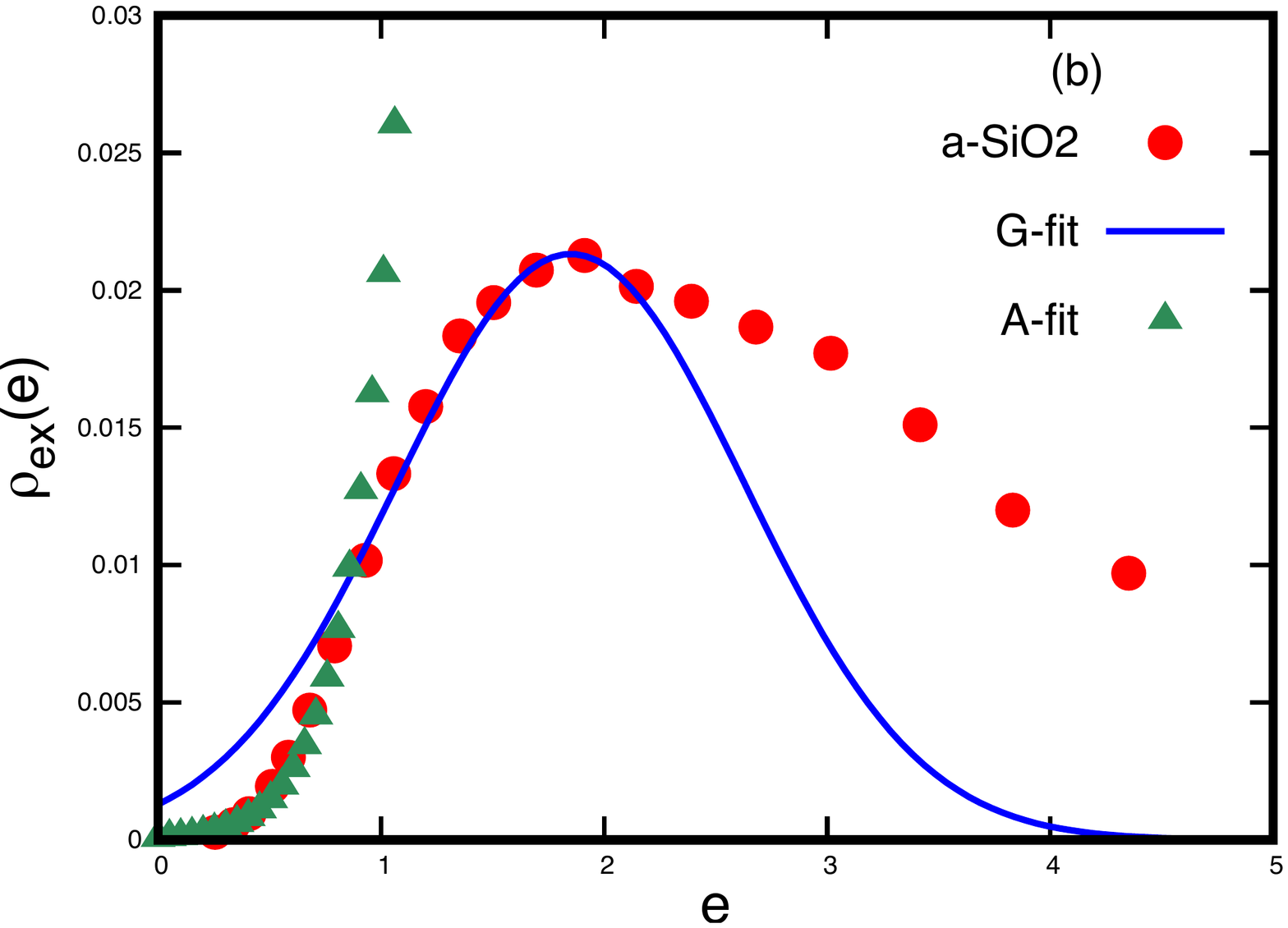}

\vspace{-1.5in}

\caption{{\bf Vibrational DOS for a-SiO$_2$:}(with $e$ in THz units)  
The experimental data here is obtained by a digital scan of the figure 12 in \cite{ya29} for reduced DOS with fits details as follows:  ''T-fit''= $2.0 {\rm Ai}(-2.85 (x-1.85)) \; (1-\Theta(x-0.7)) +{\Theta(x-0.7)\over \sqrt{1000 \pi}}  {\rm exp} \left[-{ (x-3.6)^2\over 2 (1.6)} \right] +0.004 \; x^2$, ''G-fit''=${1\over \sqrt{700 \pi}}  {\rm exp}\left[-{(x-1.85)^2\over 2 (0.61)}  \right]$ and ''A-fit''= ${\rm Ai}(x)=1.1 \; {\rm Ai} \left[-2.85 (e-1.85)\right]$. All other details are same as in figure 2.   Here again both the Gaussian as well as Airy function fitting for the case (a) is slightly different from that of (b). The ''T-fit'' in part (a) gives ${\mathcal D}_b=0.0042 \; {THz}^{-3}$ which is close to ${\mathcal D}_b= 0.0026 \; {THz}^{-3}$ given by figure 12 of \cite{ya29}. It is not clear however why the experimental excess DOS  is deviating from the Gaussian beyond $e_{gl}$ so rapidly.} 
\end{figure}

\begin{figure}[ht!]
\centering
\includegraphics[width=0.8\textwidth,height=1.\textwidth]{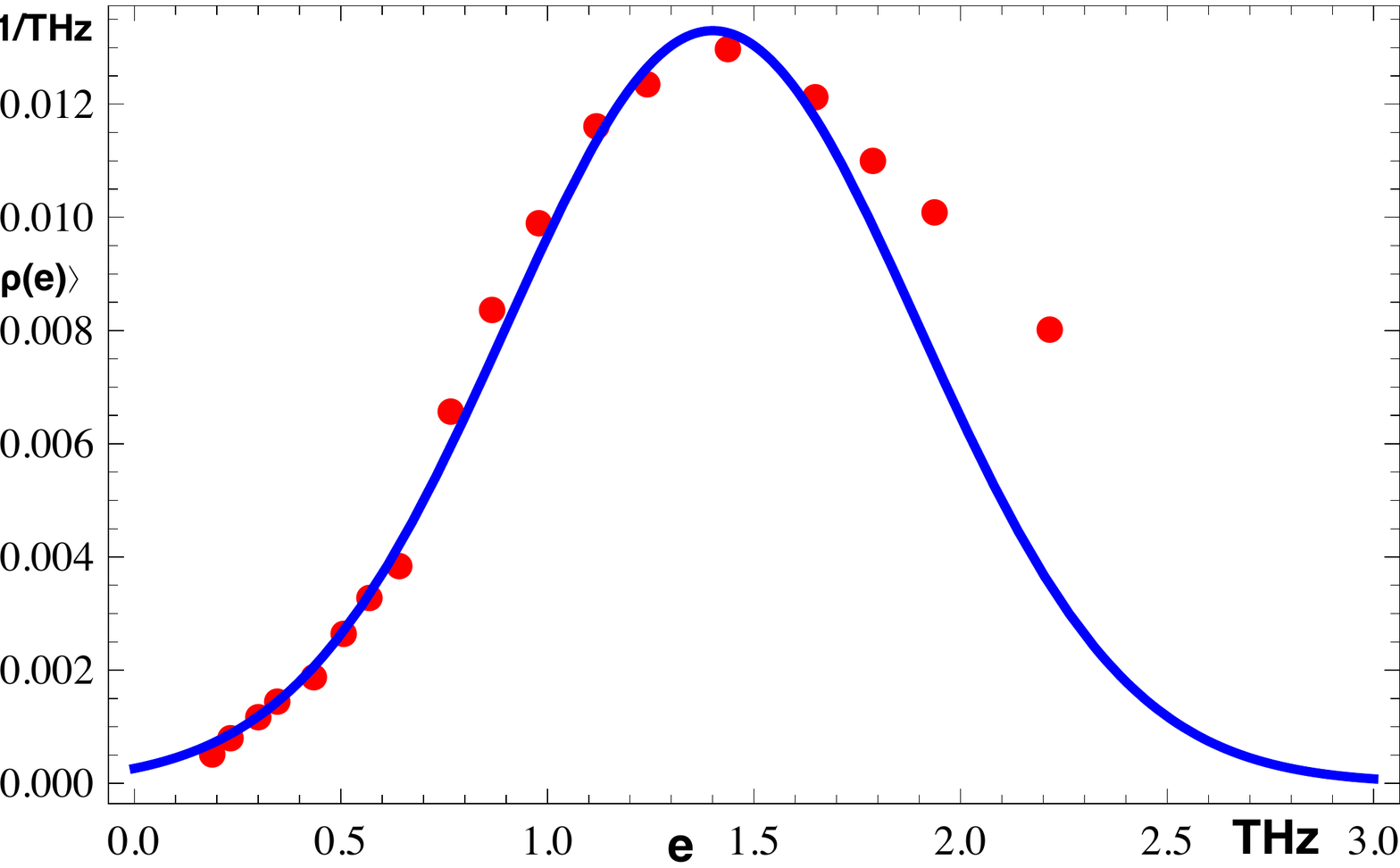}
\caption{{\bf Vibrational DOS for another SiO$_2$ sample:} (with $e$ in THz units)
A comparison of experimental excess vibrational density of states for a glass solid, taken from \cite{buch2} (adapted from figure 6 therein), with a Gaussian fit $\rho(e)={1\over \sqrt{1800 \pi }} {\rm e}^{-2 (x-1.4)^2}$; we note that the source of experimental data here is different from figure 6. 
The mean $\mu=1.4$ and variance $\sigma^2=0.5$ of the fitted Gaussian is consistent with our theoretical prediction eq.(\ref{rhoe1}). Note however, with  experimental data  normalized for a different energy range, the fitted Gaussian has a prefactor different from eq.(\ref{rhoe1})}.
\label{fig1}
\end{figure}

\begin{figure}[ht!]
\centering

\includegraphics[width=14cm,height=10cm]{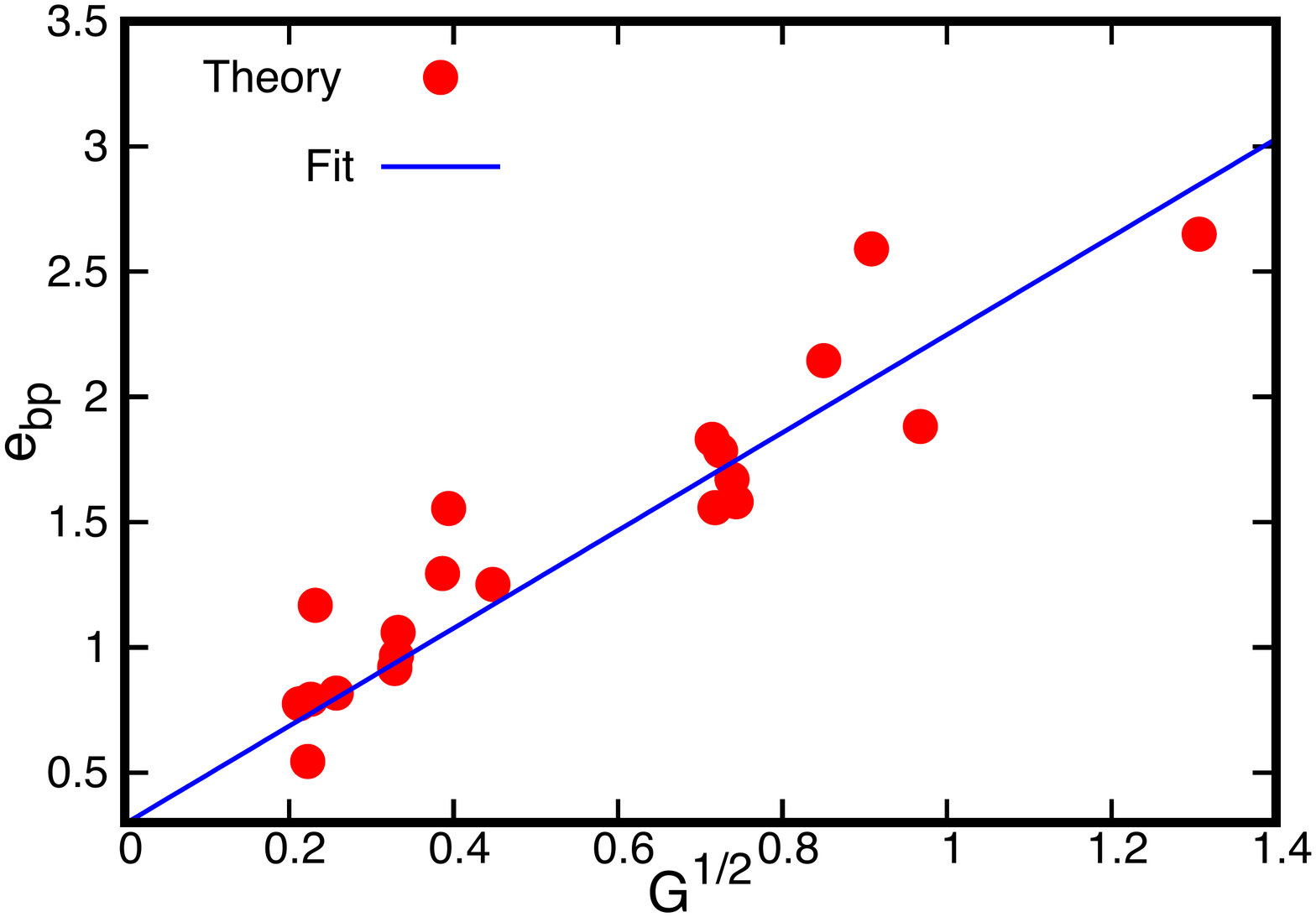}

\caption{{\bf G-dependence of boson peak frequency (with $e_{bp}$ in THz units):} 
The figure displays the dependence of the   boson peak location $e_{bp}$ on the bulk modulus $G$ for 22  glasses listed in Tables I and II. The  $G$-values for each glass are obtained   from  the $v_l, v_t$ values listed in the tables. The figures indicates a square-root dependence of $e_{bp}$ on $G$ which is consistent with the experimental observation.}
\end{figure}


\newpage

\begin{thebibliography}{10}


\bibitem{buch2}
U. Buchenau, N. Nucker and A.J. Dianoux, Phys. Rev. Lett., 53, 2316, (1984).


\bibitem{vdos}
V. K. Malinovsky, V. N. Novikov, P.P. Parashin, A.P. Solokov and M.G. Zemlyanov, Europhys. Lett., 11, 43 (1990).

\bibitem{dpr}
Due to intense interest, many research papers have been published over the years on the topic and it is not feasible to include all of them here. For example, many references on the topic can be found in {\it Dynamics of Disordered Materials II.}, edited by A.J. Dianoux, W. Petry and D. Richter (North-Holland, Amsterdam, 1993).


\bibitem{ell3}
S.R.Elliott, Europhys. Lett. 19, 201 (1992).
 


\bibitem{buch}
U. Bucheanau, Y. M. Galperin, V. Gurevich, D. Parashin, M. Ramos and H. Schober, Phys. Rev. B 46, 2798, (1992);  43, 5039, (1991).



\bibitem{spm}
V.G.Karpov, M.I.Klinger, F.N.Ignatiev, Sov. Phys. JETP 57, 439, (1983).


\bibitem{gure}
V. Gurevich, D. Parashin and H. Schrober, Phys. Rev. B, 67, 094203, (2003). 

\bibitem{psg}
D. Parashin, H. Schrober and V. Gurevich, Phys. Rev. B 76, 064206, (2007).

\bibitem{rpcv}
B. Ruffle, D.A.Parashin, E.Courtens and R. Vacher, arXiv:0711.0461

\bibitem{sdg}
W. Schirmacher, G. Diezemann and C. Ganter, Phys. Rev. Lett., 81, 136, (1998).



\bibitem{tara}
S N Taraskin and S R Elliott, Phys Rev B 61,12031, (2000); 
S N Taraskin, Y.L.Loh, G.Natrajan and S R Elliott, Phys Rev Lett 86, 1255, (2001).



\bibitem{schi2}
W. Schirmacher, Europhys. Lett. 73, 892, (2006); 
 A. Maruzzo, W. Schirmacher, A. Fratalocchi and G. Ruocco, Sci. Rep., 3, 1407, (2013).


\bibitem{grig} T.S. Grigera, V. Martin-Mayor, G. Parisi, P. Verrocchio, Nature 422
(2003) 289;
G. Parisi, J. Phys.: Condens. Matter 15 (2003) S765, and references therein.


\bibitem{srs}
W. Schirmacher, G. Ruocco and T. Scopigno, Phys. Rev. Lett. 98, 025501, (2007).

\bibitem{gual}
V. Gurarie and A. Altland, Phys. Rev. Lett. 94, 245502, (2005).

\bibitem{wyt}
M. Wyart, Europhys. Lett. 89, 64001, (2010).

\bibitem{wy1}
M. Wyart, S. R. Nagel and T A Witten, Europhys. Lett. 72, 486, (2005).

\bibitem{wy2}
M. Wyart, S. R. Nagel and T A Witten, Phys. Rev. E. 72, 051306, (2005).


\bibitem{degi}
E. DeGiuli, A. Laversanne-Finot, G. During, E. Lerner and M. Wyart, Soft Matter, 10, 5628, (2014).


\bibitem{mizu}
H. Mizuno, H. Shiba and A. Ikeda, PNAS 114, E9767, (2017). 



\bibitem{ml}
M.L.Manning and A J Liu, 109, 36002, (2015).

\bibitem{smmm}
E. Stanifer, P.K.Morse, A.A.Middleton and M.L.Manning, Phys. Rev. E, 98, 042908, (2018). 

\bibitem{bp}
Y.M.Beltukov and D.A.Parashin, {\it Physics of the solid state}, 53, 151, (2011). 

\bibitem{zac}
M. Baggioli and A. Zaccone, Phys. Rev. Research 1, 012010(R), (2020); 
M. Baggioli, R. Milkus, and A. Zaccone, Phys. Rev. E 100, 062131, (2019).

\bibitem{osln}
C S Ohern, L E Silbert, A J Liu and S R Nagel, Phys. Rev. E, 68, 011306, (2003).


\bibitem{lw} V. Lubchenko and P. G. Wolynes, Proc. Natl. Acad. Sci. USA 100, 1515 (2003).

\bibitem{du}
E. Duval, A. Boukenter, T. Achibat,  J. Phys. Condens. Matter 2, 10227, (1990). 

\bibitem{mns}
V K Malinovsky, V N Novikov and A P sokolov, Phys. Lett. A, 153, 63, (1991).


\bibitem{sksq}
A. P. Sokolov, A. Kisliuk, M. Soltwisch, and D. Quitmann Phys. Rev. Lett. 69, 1540 (1992).

\bibitem{fpuz} 
S. Franz, G. Parisi, P. Urbani and F. Zamponi, PNAS, 112, 14359, (2015).

\bibitem{bb1} P. Shukla, arXiv:2008.12960.
\bibitem{bb2} P. Shukla, arXiv:2009.00556.
\bibitem{bb3} P. Shukla, arXiv:2101.00492


\bibitem{mg}
G. Monaco and V. M. Giordano, PNAS.0808965106.



\bibitem{isra}
J. Israelachvili, Chapter 11, {\it Intermolecular and Surface Forces}, 3rd ed. Academic Press, (2011).

\bibitem{ajs}
A.J.Stone, {\it The theory of intermolecular forces}, Oxford scholarship online, Oxford university Press, U.K. 2015.


\bibitem{arg}
C. Argento, A. Jagota, W.C. Carter, J. Mech. Phys. Solids 45, 1161, (1997).

\bibitem{meek}
R.M. Meeking, J. Colloid Interface Sci. 199, 187, (1998).

\bibitem{lhe}
L.H.He, J. Mech. Phys. Solids 61, 1377, (2013).


\bibitem{vl}
D. Vural and A.J.Leggett, J. Non crystalline solids, 357, 19, 3528, (2011). 



\bibitem{dl}
Z. Dee and A. J. Leggett, arXiv: 1510:05528v1.

\bibitem{lg1}
A. J. Leggett and D. Vural, J. Phys. Chem. B., 42,117, (2013).


\bibitem{jl}
J. Joffrin and A. Levelut, Jou.de. Physique, 36, 811, (1975).

\bibitem{paras}
D. A. Parashin, Phys. Rev. B, 49, 9400, (1994). 



\bibitem{hjs}
H-J Stockmann, {\bf Quantum Chaos: an introduction}, Cambridge univ. Press (1999) (see page 79). 


\bibitem{issr}
 L. Isserlis, Biometrika, 12, 134, (1918).


\bibitem{bray}
G. J. Rodgers and A. J. Bray, Phys. Rev. B, 37, 3557, (1998). 

\bibitem{khor}
A. Khorunzhy and G. J. Rodgers, J. Math. Phys. 38, 3300 (1997).

\bibitem{thou}
R.C Jones, J M Kosterlitz and D J Thouless,  J. Phys. A: Math. Gen., 11, 3,1978.


\bibitem{lern}
E. Lerner, G. Düring, and E. Bouchbinder, Phys. Rev. Lett., 117, 035501, (2016).


\bibitem{chum}
A. I. Chumakov et al., Phys. Rev. Lett. 106, 225501 (2011).

\bibitem{wang}
Y. Wang, L. Hong, Y. Wang, W. Schirmacher and J. Zhang, Phys. Rev. B 98, 174207 (2018). 
 
 \bibitem{nie}Y. Nie, H. Tong, J. Liu, M. Zu, N. Xu, Front. Phys. 12, 126301 (2017)

\bibitem{brody}
T.A.Brody, J.Flores, J.B.French, P.A.Mello, A. Pandey and S.S.M Wong, Rev. Mod. Phys., 53, 1981.



\bibitem{stol}
R.H. Stolen, Phys.Chem.Glasses, 11, 83,  (1970).

\bibitem{nema}
R.J. Nemanich, Phys.Rev.B, 16, 1665, (1977).



\bibitem{novi}
V.N. Novikov, and A.P. Sokolov, Sol.State Comm., 77, 243, (1991).



\bibitem{prat}
 J.L. Prat, F. Terki, and J. Pelous, Phys.Rev.Lett., 77, 755, (1996).

\bibitem{kgb}
K.G. Breitschwerdt, and S. Gut, Proc. 12 Intern. Conf. on acoustic, Toronto, G2-6, (1986).


\bibitem{ryz}
V A Ryzhov,  Phys Astron Int J. 3, 123, (2019).




\bibitem{pala}
D. A. Parashin and C. Laermans, Phys. Rev. B 63,  132203, (2001).

\bibitem{pnes}
S. Perticaroli, J. D. Nickels, G. Ehlers and A. P. Sokolov, Biophysical Journal,  106, 2667, (2014).

\bibitem{yanno}
S.N. Yannopoulos, K.S. Andrikopoulos, G. Ruocco, J. Non-Cryst. Sol. 352, 4541, (2006).

\bibitem{ya21} A. Tolle, H. Zimmermann, F. Fujara, W. Petry, W. Schmidt, H.
Schober, J. Wuttke, Eur. Phys. J. B 16 , 73, (2000).

\bibitem{ya25} W.A. Phillips, U. Buchenau, N. Nucker, A. J.Dianoux, W. Petry, Phys. Rev. Lett. 63, 2381, (1989).

\bibitem{ya26} D. Engberg, A. Wischnewski, U. Buchenau, L. Borjesson, A.J.
Dianoux, A.P. Sokolov, L.M. Torell, Phys. Rev. B 58, 9087, (1998).

\bibitem{ya27} R. Zorn, A. Arbe, J. Colmenero, B. Frick, D. Richter, U. Buchenau, Phys. Rev. E 52, 781, (1995).

\bibitem{ya28}  J. Wuttke, W. Petry, G. Goddens, F. Fujara, Phys. Rev. E 52, 4026, (1995).

\bibitem{ya29} A. Wischnewski, U. Buchenau, A.J. Dianoux, W. A. Kamitakahara, J. L. Zarestky, Phys. Rev. B 57, 2663, (1998).

\bibitem{mmb}
H. Mizuno, S. Mossa and J.-L. Barrat, arXiv:1308.5135, (2013).

\bibitem{hsgms}
R. P. Hermann, R. JIn, W. Schweika, F. Grandjean, D. Mandrus, B. C. Sales, G.J. Long, Phys. Rev. Lett. 90, 135505, (2003).



\bibitem{pohl}
R.O.Pohl, X.Liu and E.Thompson, Rev. Mod. Phys. 74, 991, (2002). 

\bibitem{mb}
J.F. Berret and M. Meissner, Z. Phys. B-Condensed Matter 70, 65, (1988).

\bibitem{bb5}
P. Shukla, to be submitted.

\bibitem{hw}
Y. Higashigaki and C.H. wang, J. Chem. Phys. 74, 3175, (1981).

\bibitem{lya}
A.G.Lyapin, RSC ADV,7 33278, (2017).



\end{thebibliography}
\end{document}